\documentclass[aip,jcp,numerical,reprint]{revtex4-2}

\usepackage{amsmath}
\usepackage{epstopdf}
\usepackage{graphicx}
\usepackage{dcolumn}
\usepackage{bm}
\usepackage{braket}
\usepackage{gensymb}
\usepackage{array}
\newcolumntype{P}[1]{>{\centering\arraybackslash}p{#1}}
\newcolumntype{M}[1]{>{\centering\arraybackslash}m{#1}}

\newcommand{\eps}{$\varepsilon$}
\newcommand{\tinter}{$t_{\textrm{inter}}$}
\newcommand{\tintra}{$t_{\textrm{intra}}$}

\newcommand{\buplus}[1]{${#1}^1B_u^+$}
\newcommand{\TTTsep}{$^3|T \cdots T\rangle_{1-2}$}
\newcommand{\QTTsep}{$^5|T \cdots T\rangle_{1-2}$}
\newcommand{\TTsep}{$^1|T \cdots T\rangle_{1-2}$}
\newcommand{\TT}[1]{$^1|TT\rangle_{#1}$}
\newcommand{\T}[1]{$|T\rangle_{#1}$}

\newcommand{\buminus}[1]{${#1}^1B_u^-$}

\newcommand{\agminus}[1]{${#1}^1A_g^-$}

\begin{document}

\title{Theory of Singlet Fission in Carotenoid Dimers}

\author{William Barford}
\email{william.barford@chem.ox.ac.uk}
\affiliation{Department of Chemistry, Physical and Theoretical Chemistry Laboratory, University of Oxford, Oxford, OX1 3QZ, United Kingdom}
\affiliation{Balliol College, University of Oxford, Oxford, OX1 3BJ, United Kingdom}

\author{Cameron A.\ Chambers}
\affiliation{Department of Chemistry, Physical and Theoretical Chemistry Laboratory, University of Oxford, Oxford, OX1 3QZ, United Kingdom}
\affiliation{Lincoln College, University of Oxford, Oxford, OX1 3DR, United Kingdom}

\date{\today}

\begin{abstract}
We develop a theory of singlet fission in carotenoid dimers. Following photoexcitation of the `bright' state (i.e., a singlet electron-hole pair) in a single carotenoid, the first step in the singlet fission process is ultrafast  intramolecular conversion into the highly-correlated `dark' (or $2A_g$) state. This state has both entangled singlet triplet-pair and charge-transfer character. Our theory is predicated on the assumption that it is the singlet triplet-pair component of the `dark' state that undergoes bimolecular singlet fission.
We use valence bond theory to develop a minimal two-chain model of the triplet-pair states.  The single and double chain triplet-pair spectrum is described, as this helps explain  the dynamics and the equilibrated populations. We simulate the dynamics of the initial entangled pair state using the quantum Liouville equation,
 including both spin-conserving and spin-nonconserving dephasing processes. By computing the intrachain and interchain singlet, triplet and quintet triplet-pair populations, we show that singlet fission depends critically on the interchain coupling and the driving potential (that determines endothermic versus exothermic fission). We also show that the Horodecki pair-entanglement provides a good metric for singlet fission.
\end{abstract}

\maketitle


\section{Introduction}\label{Se:1}

Singlet fission is a photophysical process whereby a photoexcited electron-hole pair initially forms a pair of correlated electron-hole pairs that eventually dissociate and decohere into separate electron-hole pairs\cite{Smith2010,Casanova2018,Musser19,Sanders19,Zhu2019}. Owing to optical selection rules, for a singlet groundstate the initial electron-hole pair is also a spin-singlet state. The strong electronic correlations present in low-dimensional conjugated molecules means that the exchange energy between the singlet and triplet electron-hole pairs is so large that the singlet electron-hole energy is approximately twice the triplet electron-hole energy\cite{Book}.
Consequently, the subsequent spin-conserving process of the formation of two electron-hole pairs  implies that this an  entangled pair of two  triplets in an overall singlet state.
In contrast with other researchers\cite{Zhu2019}, in this paper we  explicitly define singlet fission as the process by which a singlet electron-hole pair interconverts to an entangled pair of triplets, which  ultimately dissociates and spin-decoheres into two separate, uncorrelated triplets. A quantitative definition will be given in Section \ref{Se:5.3}.

Singlet fission has been widely investigated in acene molecules. The work on acenes is partly motivated by the observation that the creation of a pair of electron-hole pairs (by a photon of twice the excitation energy of each pair) in tandem with a material that produces a single electron-hole pair at the same energy (by a photon of half the energy for the singlet fission material) has the potential to exceed the Shockley-Queisser limit\cite{Ross1980,Hanna2006}.

In acenes it is widely accepted that the first step in the overall process (i.e., the conversion of one electron-hole pair into two correlated electron-hole pairs) is a bimolecular process\cite{Reichman2013a,Reichman2013b,Mazumdar2015,Casanova2018}. In this case, via a two-electron process, the initial electron-hole pair created on one molecule forms a pair of electron-hole pairs delocalized over two molecules.

The study of singlet fission in carotenoids and polyenes is at a less mature stage than for acenes\cite{Zimmerman2018,Musser19,Ghosh2022}. In long polyene-type systems there is evidence of triplet formation via singlet fission on single chains\cite{Kraabel98,Lanzani99,Musser13}. For shorter polyenes, e.g., carotenoids, however, singlet fission appears to only occur in aggregates and dimers\cite{Wang11,Zhu2014,Musser15,Clark2023}. Owing to both strong electron-electron interactions and electron-nuclear coupling, the excited states of polyenes and carotenoids are more complex than those of acenes. In particular, the well-known `bright' to `dark' state internal conversion\cite{Manawadu2022,Manawadu2023a,Manawadu2023b} leads to the formation of a highly correlated low-energy state (usually labelled $2A_g$ or S$_1$).

\begin{figure}[h]

\includegraphics[width=0.9\linewidth]{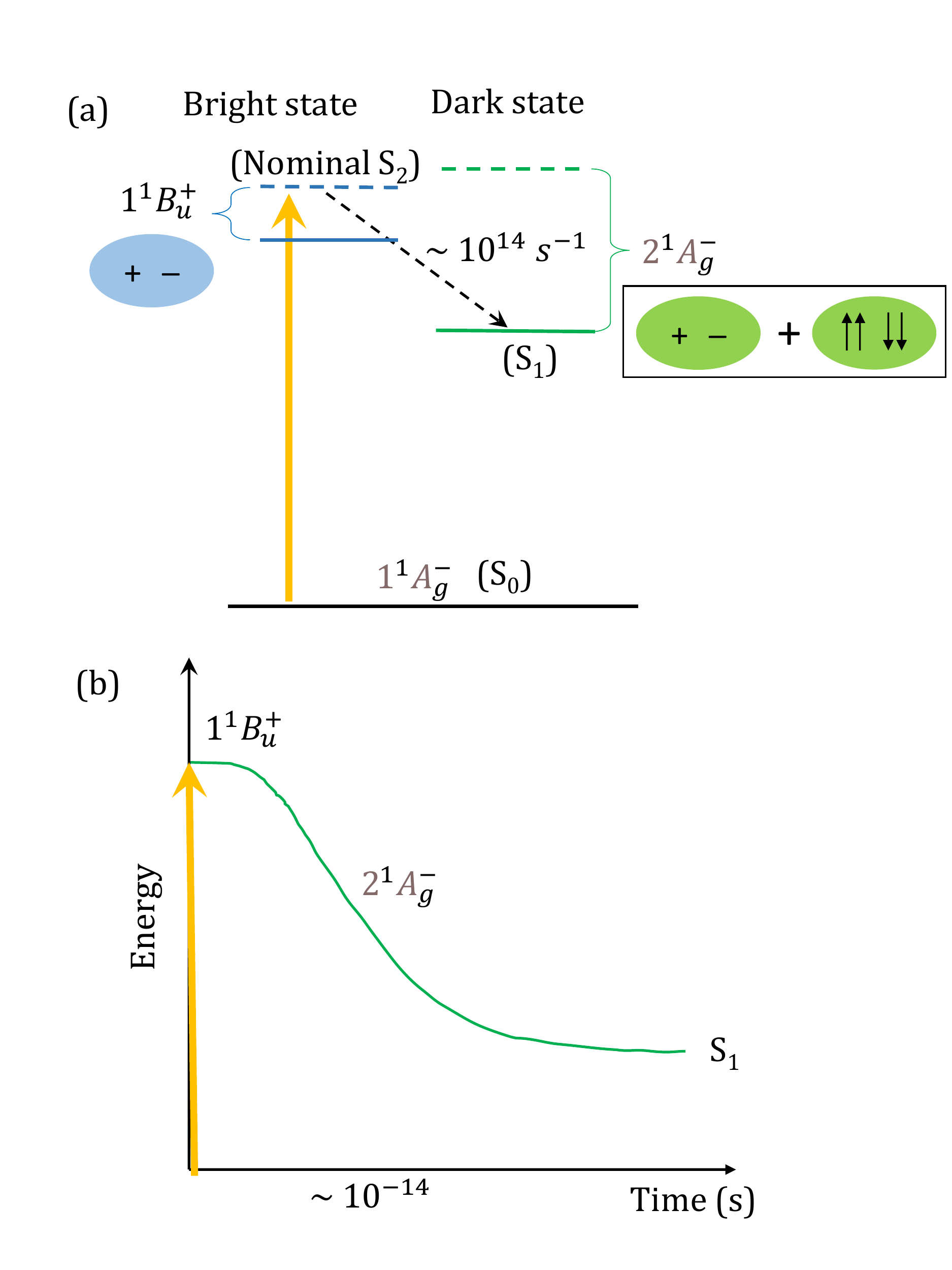}
\caption{Schematic diagrams illustrating internal conversion between the excited `bright' and `dark' states of a carotenoid.
 (a) Diabatic state representation: $1^1 B_u^+$ (blue) and $2^1 A_g^-$ (green).
  The dashed (solid) horizontal lines represent the vertical (relaxed) energies of the states. The approximate rate constant is also indicated.
  (b) Adiabatic state representation, showing $S_1$ evolve adiabatically from  the predominately \buplus{1} state (i.e., a Frenkel exciton) to the predominately \agminus{2} state (i.e., a linear superposition of a singlet triplet-pair and an odd-parity charge-transfer exciton).
  (Internal conversion from the  \buplus{1}   to  \agminus{2} states   via the intermediate \buminus{1} state is illustrated in Fig.\ 4 of ref\cite{Manawadu2023a}.)}
\label{Fi:17}
\end{figure}

One possible `bright' to `dark' state internal conversion process in carotenoids is illustrated in Fig.\ \ref{Fi:17}. This shows an energy level crossing between the diabatic \buplus{1} and \agminus{2} states, occurring  within 10 fs of photoexcitation as a consequence of strong electron-nuclear coupling\cite{Manawadu2023a}. In an adiabatic representation\footnote{The  `diabatic' \buplus{1} and \agminus{2} states are eigenstates of a model Hamiltonian which assumes that carotenoids posses $C_2$ and particle-hole symmetry. In contrast, the `adiabatic' $S_1$ and $S_2$ states  are eigenstates of a model Hamiltonian which assumes that carotenoids do not posses  $C_2$ and particle-hole symmetry\cite{Manawadu2023a}.} the S$_1$ state, initially  predominately the \buplus{1} state, exhibits an \emph{avoided crossing} with the  S$_2$ state, initially  predominately the \agminus{2} state, such that after 10 fs S$_1$ becomes predominately the \agminus{2} state. However, this is not the only possible `bright' to `dark' state internal conversion process\cite{Frank1997,Kosumi2006,Manawadu2022,Manawadu2023a}. This is because the \agminus{2} state is just the lowest-energy member of a band (or family) of `$2A_g$' states\cite{Valentine20}, namely, states of the same fundamental excitation with different center-of-mass pseudo momenta, i.e., \agminus{2}, \buminus{1}, \agminus{3}, etc. Thus, another internal conversion process from the  \buplus{1} to \agminus{2} states is via the intermediate \buminus{1} state. These two processes are described in detail in ref\cite{Manawadu2022,Manawadu2023a,Manawadu2023b}.

We now  turn to a more detailed discussion of the `$2A_g$' family of states. As shown in refs\cite{Valentine20,Barford2022c,Manawadu2023a}, these states are a linear superposition of a singlet triplet-pair and an odd-parity charge-transfer exciton (as illustrated schematically in Fig.\ \ref{Fi:17}(a)). The hybridization between these two components causes a strong triplet-triplet attraction\cite{Barford2022c}. It also implies that the energy of the \agminus{2} state is ca.\ 0.4 eV lower than the energy of a pair of noninteracting triplets on a chain of the same length\cite{Valentine20}. Thus, intrachain singlet fission from the \agminus{2} state is a strongly endothermic process (although it becomes less exothermic and potentially endothermic from higher energy members of the `$2A_g$' family\cite{Valentine20,Manawadu2022} - a point that we return to in the Conclusions).

As already mentioned, singlet fission in acenes is an intrinsically bimolecular process. However, the  triplet-pair character of the \agminus{2} state of polyenes suggests that for these systems the initial step of the formation of an entangled triplet-pair is  an intrinsically unimolecular process. In this paper we focus on singlet fission from the \agminus{2} state. Owing to strong endothermic intrachain dissociation for this state, we propose that the second step towards energetically favourable singlet fission is a bimolecular mechanism. In ref\cite{Manawadu2023a} a possible mechanism  of bimolecular exothermic dissociation was suggested. This mechanism relies on the additional vibrational and torsional reorganization energies that a pair of individual triplets on separate chains gain over a pair of bound triplets on the same chain. It thus assumes that carotenoids are in a twisted configuration in their ground state and planarize in their excited states.

In this paper we introduce a minimal model of triplet-pair dissociation and decoherence in carotenoid dimers. We start by assuming that the \agminus{2} state formed via internal conversion on a single chain is entirely composed of a correlated singlet triplet-pair. The triplets on a single chain interact with one another and delocalize along the chain via a superexchange mechanism. A triplet may also hop onto a neighboring chain, thereby losing its triplet-pair interaction. However, by doing so it gains an off-set (or driving) energy from the additional reorganization energies discussed in the previous paragraph.

In order to understand the singlet fission process we compute various metrics. We distinguish between bound, intrachain singlet triplet-pairs and noninteracting, interchain singlet, triplet and quintet triplet-pairs, and we compute their populations.
Since the dynamics and equilibrium properties are ultimately determined by the energy spectrum of the quantum system, we discuss the single and double chain spectra in some detail.
A particularly useful attribute is the Horodecki pair-entanglement\cite{Horodecki2009,Marcus20}: we show that this vanishes at complete singlet fission. We model the dynamics of the open  quantum system using the quantum Liouville equation in which both spin-conserving and spin interconversion dephasing processes are included.

The plan of this paper is as follows. In the next section we introduce and motivate our model (a fuller justification is given in the Appendix). Section III describes the singlet, triplet and quintet triplet-pair spectrum of the single and coupled chain Hamiltonians. Section IV  describes our computational methodology and introduces the key metrics by which we investigate the singlet fission process. We present and discuss both our dynamical and equilibrium results in Section V. We conclude and outline proposals for future work in Section VI.


\section{Model of Triplet-Pair States}\label{Se:2}

\subsection{The Triplet-Pair Basis}

\begin{figure}
\includegraphics[width=0.9\linewidth]{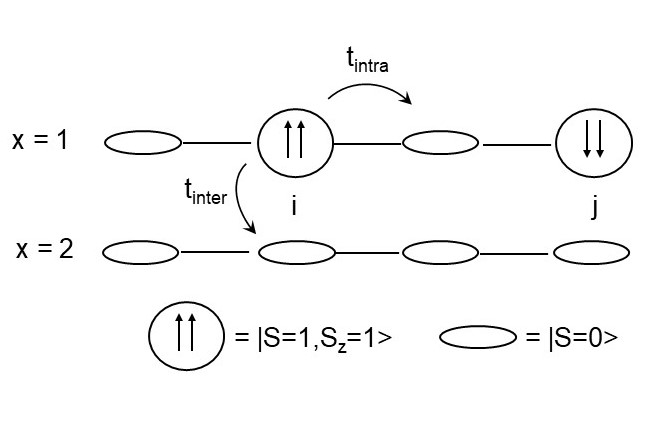}
\caption{A schematic illustration of a carotenoid dimer with a triplet-pair on chain $\times=1$, with triplets on ethylene dimers $i$ and $j$.
Chain $\times=2$ is in its ground state.
\tintra\ and \tinter\ are the hopping matrix elements between neighboring intrachain and interchain ethylene dimers, respectively.}
\label{Fi:18}
\end{figure}

Our model for the triplet-pair states of carotenoids and polyenes is based on the valence bond model of strongly correlated systems\cite{Coulson}, as illustrated in Fig.\ \ref{Fi:18}. The ground state is shown by the chain labeled $\times =2$. Each $p_z$ orbital is singly occupied, while both electrons on an ethylene dimer form a singlet bond. A triplet excitation on the $i$th dimer is denoted as $|0;i\rangle$ or $|\pm 1;i\rangle$, for the $M_s=0$ or $M_s=\pm 1$ spin-projections, respectively. A pair of triplet excitations on   dimers $i$ and $j$ is shown for the chain labeled $\times =1$ in Fig.\ \ref{Fi:18}.

As described in the Introduction,  we assume that within tens of fs the photoexcited Frenkel exciton undergoes internal conversion to a singlet triplet-pair state on  a single chain (e.g., $\times =1$). The initial entangled pair is thus,
\begin{equation}\label{Eq:01}
 {^1|\Phi(t=0)\rangle} = \sum_{ij \in \times = 1} \Phi_{ij}(t=0) {^1|i,j\rangle},
\end{equation}
where the singlet triplet-pair basis state is
\begin{equation}\label{Eq:02}
  ^1|i,j\rangle = \frac{1}{\sqrt{3}}\left( |1;i\rangle|-1;j\rangle - |0;i\rangle|0;j\rangle + |-1;i\rangle|1;j\rangle \right).
\end{equation}
For $ {^1|\Phi(t=0)\rangle}$, $i$ and $j$ label dimers in chain $\times =1$, but for a general eigenstate of the full two-chain Hamiltonian introduced in the next section $i$ and $j$ can label dimers on separate chains.

A pair of triplets within the valence bond basis are coupled to form an overall singlet, triplet or quintet state\cite{}. In this work we investigate the role of transverse spin-dephasing which connects the $S_z=0$ components of each total spin.
The $S_z = 0$ components of the triplet and quintet triplet-pair bases are,
\begin{equation}\label{}
  ^3|i,j\rangle = \frac{1}{\sqrt{2}}\left( |1;i\rangle|-1;j\rangle - |-1;i\rangle|1;j\rangle \right),
\end{equation}
and
\begin{equation}\label{}
  ^5|i,j\rangle = \frac{1}{\sqrt{6}}\left( |1;i\rangle|-1;j\rangle + 2|0;i\rangle|0;j\rangle + |-1;i\rangle|1;j\rangle \right),
\end{equation}
respectively.

\vspace{0.5cm}

\subsection{The Two-Chain Hamiltonian}

Assuming  the valence bond model approximation, the low-energy physics of carotenoids and polyenes are then described by the spin-1/2 Heisenberg antiferromagnet model. As described in more detail in Appendix A, within the reduced triplet-pair basis introduced above, the Heisenberg antiferromagnet  for a pair of chains reduces to the following Hamiltonian,
\begin{equation}\label{Eq:400}
   \hat{H} = \sum_{\times=1,2} \hat{H}_{\textrm{single}}^{\times}  + \hat{H}_{\textrm{double}} +  \hat{H}_{\textrm{inter}}.
\end{equation}

$\hat{H}_{\textrm{single}}^{\times}$ is the intrachain Hamiltonian for a pair of triplets on the same chain, $\times$:
\begin{widetext}
\begin{eqnarray}\label{Eq:401}
  \hat{H}_{\textrm{single}}^{\times=1,2} =
 2E_{\textrm{T}}\sum_{i, j>i \in \times}  \left|i,j\rangle \langle i,j\right|
 +  t_{\textrm{intra}}\sum_{i \ne j \in \times}\left(\left|i \pm 1,j\rangle \langle i,j\right| + \textrm{H.C.} \right)
 - V \sum_{i \in \times}  \left|i,i+1\rangle \langle i,i+1\right|.
 \nonumber \\
\end{eqnarray}
\end{widetext}
The first term on the right-hand-side describes the excitation of a pair of triplets on dimers $i$ and $j$ on the same chain, while the second term describes the hopping of triplets between neighboring dimers on the same chain. The final term describes the spin-dependent exchange interaction between a pair of triplets on neighboring dimers. The origin of this interaction is explained in Appendix A. This interaction may also be written as\cite{Kollmar1993}
\begin{equation}\label{}
{J}\hat{\textbf{S}}_i^{(1)} \cdot \hat{{\textbf{S}}}_{i+1}^{(1)},
\end{equation}
where $\hat{\textbf{S}}^{(1)}$ is the  spin-1 operator. Thus, the singlet and triplet triplet-pairs experience a nearest-neighbour attraction, $V_S = +2J$ and $V_T = +J$, respectively, while the quintet triplet-pair experiences a nearest-neighbour repulsion $V_Q = -J$.

$\hat{H}_{\textrm{double}}$ is the intrachain Hamiltonian for a pair of triplets on separate chains:
\begin{widetext}
\begin{eqnarray}\label{Eq:402}
  \hat{H}_{\textrm{double}} =
2(E_{\textrm{T}}-\varepsilon)\sum_{i \in \times=1} \sum_{j \in \times=2} \left|i,j\rangle \langle i,j\right|
+ t_{\textrm{intra}}\sum_{i \in \times=1} \sum_{j \in \times=2} \left[ \left(\left|i \pm 1,j\rangle \langle i,j\right| + \textrm{H.C.} \right)+
   \left(\left|i,j \pm 1\rangle \langle i,j\right| + \textrm{H.C.} \right) \right].
  \nonumber\\
\end{eqnarray}
\end{widetext}
The first term on the right-hand-side describes the excitation of a pair of triplets on dimers $i$ and $j$ on different chains. \eps\ is the offset energy which each triplet gains because of the additional reorganization energy that each separate triplet gains over a bound pair on the same chain.
The second and third terms describe the hopping of triplets between neighboring dimers on the same chain.

Finally, $\hat{H}_{\textrm{inter}}$ is the interchain Hamiltonian which couples both chains:
\begin{widetext}
\begin{eqnarray}\label{Eq:403}
  \hat{H}_{\textrm{inter}} =
  t_{\textrm{inter}} \sum_{\times = 1}^2\sum_{i_{\times}} \sum_{j_{\times}>i_{\times}}
\left[
\left(
\left|i_{\times},j_{\times}\rangle \langle i_{\bar{\times}},j_{{\times}}\right|
+ \textrm{H.C.}
\right)  +
\left(
\left|i_{\times},j_{\times}\rangle \langle i_{{\times}},j_{\bar{\times}}\right|
+ \textrm{H.C.}
\right)
\right],
\end{eqnarray}
\end{widetext}
where $i_{\times}$ and $i_{\bar{\times}}$ mean the $i$th dimer on opposite chains.
This describes the hopping of triplets between nearest neighbor dimers of both chains.

\section{The Triplet-Pair States And Spectrum}\label{Se:3}

\subsection{The Single-Chain Spectrum}\label{Se:3.1}

\begin{figure}
\includegraphics[width=0.8\linewidth]{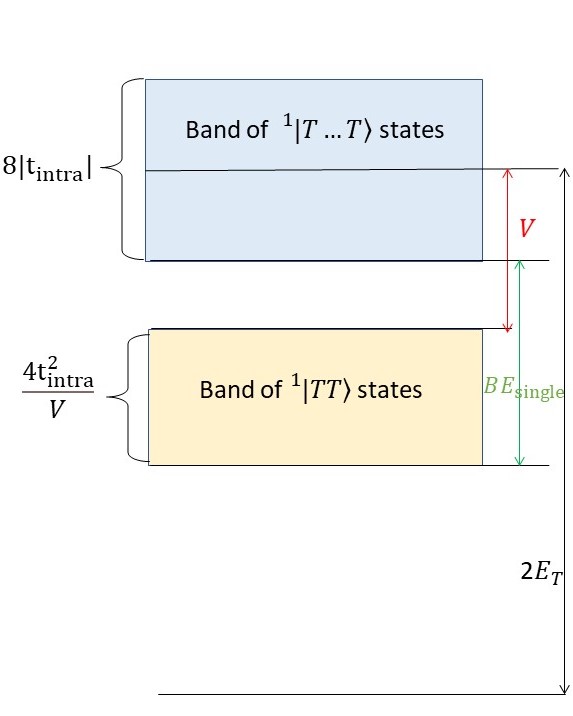}
\caption{The spectrum of the intrachain singlet triplet-pair states, as determined by eqn (\ref{Eq:06}) and eqn (\ref{Eq:05}).
$^1|TT\rangle$ are bound intrachain singlet triplet-pairs, which form the `$2A_g$' family of states (i.e., $2^1A_g^-, 1^1B_u^-, 3^1A_g^-, \cdots$). $^1|T \cdots T\rangle$ are unbound intrachain singlet triplet-pairs.
The triplet-pairs are bound if $V> V_c = 2|t_{\textrm{intra}}|$.
(An equivalent description applies to triplet triplet-pairs. In contrast, the quintet triplet-pairs are antibound.)
Compare to the inset of Fig.\ \ref{Fi:16}, which shows the energy levels for the triplet-pairs on a pair of ethylene dimers.
}
\label{Fi:14}
\end{figure}

The solution of the two-particle single chain Hamiltonian  $\hat{H}_{\textrm{single}}$, eqn (\ref{Eq:401}), is well-known\cite{Mattis1988,Gallagher1997,Gebhard1997}.
In a  chain with periodic boundary conditions there is a triplet-pair correlated state, which we denote as $|TT\rangle$, which forms a band with energy,
\begin{equation}\label{Eq:06}
E_k^{\textrm{bound}} =  (2E_T -V) - \frac{4t_{\textrm{intra}}^2}{V}\cos^2 (k/2).
\end{equation}
Here, $k$ is the dimensionless wavevector, satisfying $-\pi \le k \le \pi$.
Similarly, the energy of a pair of free, noninteracting triplets, which  we denote as $|T \cdots T\rangle$, is,
\begin{equation}\label{Eq:05}
E_k^{\textrm{free}} =  2E_T - 4t_{\textrm{intra}}\cos (k).
\end{equation}
Thus, the binding energy of the triplet-pair correlated state on a single chain is
\begin{eqnarray}\label{Eq:B7}
\textrm{BE}_{\textrm{single}} = &&  (E_{k=0}^{\textrm{free}}-E_{k=0}^{\textrm{bound}})
\nonumber\\
= && V +\frac{4 t_{\textrm{intra}}^2}{V} -   4|t_{\textrm{intra}}|.
\end{eqnarray}
This result implies that the triplet-pair is bound if
$V$ exceeds the critical value, $V_c = 2|t_{\textrm{intra}}|$.

These predictions remain qualitatively correct for linear chains with open boundary conditions, although  $V_c$ decreases for shorter chains.

Figure \ref{Fi:14} illustrates the single-chain spectrum for the singlet triplet-pair. Assuming that $V_S > V_c$, the band of bound states, labeled \TT, form the `$2A_g$' family of states (i.e., $2^1A_g^-, 1^1B_u^-, 3^1A_g^-, \cdots$). The noninteracting, but spin-correlated pairs form the $^1|T \cdots T\rangle$ band.

Owing to  the large gap between the lowest \TT{} state and the band of $^1|T \cdots T\rangle$ state, we find that the latter are never populated and so we no longer consider them.


\subsection{The Double-Chain Spectrum}\label{Se:3.2}

We define \TTsep\ as a noninteracting, interchain singlet triplet-pair eigenstate of the double chain Hamiltonian  $\hat{H}_{\textrm{double}}$, eqn (\ref{Eq:402}).

For  finite-length and open uncoupled chains, the energy to dissociate a singlet triplet-pair on a single chain (i.e., \TT{}) into two noninteracting triplets on separate chains (i.e., \TTsep) is defined by
\begin{equation}\label{Eq:BE}
  \textrm{BE} =   E_{\textrm{double}}(\varepsilon=0)-E_{\textrm{single}}(V>V_c),
\end{equation}
where $E_{\textrm{single}}$ and $E_{\textrm{double}}$ are the lowest singlet energies of $\hat{H}_{\textrm{single}}$ and $\hat{H}_{\textrm{double}}$, respectively.
This binding energy is  smaller than the binding energy on a single chain because of the quantum deconfinement of single triplets on separate chains. For example, when $V_S = 2.8$\tintra\ and for 20 C-sites on each chain, BE = 0.3272\tintra, whereas $\textrm{BE}_{\textrm{single}}=$0.5637\tintra.

The triplet and quintet triplet-pair eigenstates of  $\hat{H}_{\textrm{double}}$ are the noninteracting, interchain  pairs, denoted as \TTTsep\ and \QTTsep, respectively. These state are degenerate with \TTsep.

\subsection{The Full Two-Chain Spectrum}\label{Se:3.3}

We now discuss the spectrum of the full two-chain Hamiltonian, given by eqn (\ref{Eq:400}).
Since single triplets on separate chains each gain an offset potential energy of \eps, bound, intrachain singlet triplet-pairs and  noninteracting, interchain triplet-pairs are degenerate when 2\eps = BE. We will denote this as the `degeneracy point'. When \tinter = 0, for \eps\ $<$ BE/2 the lowest singlet eigenstate is a linear superposition of \TT{1} and \TT{2}. In this regime
singlet fission is endothermic. Conversely  for \eps\ $>$ BE/2 the lowest singlet eigenstate is \TTsep\ and singlet fission is exothermic  when \tinter = 0.

Coupling the two chains via $\hat{H}_{\textrm{inter}}$ causes \TT{} and \TTsep\ to hybridize. In general, the singlet triplet-pair eigenstates of the full two-chain Hamiltonian are a linear combination of the intra and interchain pair states, i.e.,
\begin{equation}\label{Eq:1140}
  ^1|\Psi\rangle = \frac{a}{\sqrt{2}}\left( ^1|TT\rangle_{1} \pm   ^1|TT\rangle_{2}\right) + b ^1|T \cdots T\rangle_{1-2},
\end{equation}
(where we have ignored the pseudomomentum quantum number).
The lowest energy singlet eigenstate is predominately composed of \TT{1} and \TT{2} in the endothermic regime (i.e., $|a|^2 > |b|^2$) and  predominately composed of \TTsep\ in the exothermic regime (i.e., $|a|^2 < |b|^2$).

As an example,  consider the  low-energy singlet triplet-pair spectrum at the degeneracy point, \eps\ $=$ BE/2.
Denoting the degenerate trio of (diabatic) singlet triplet-pair states as \TT{1} $\equiv |1\rangle$,  \TTsep $ \equiv |2\rangle$, and  \TT{2} $ \equiv |3\rangle$, the interchain coupling causes these states to hybridize to form the eigenstates $^1|\Psi_{j} \rangle = \frac{1}{2} \sum_n \sin\left(\pi j n/4\right) |n \rangle$ (where $j=1,2,3$).
In particular, the bonding eigenstate  $^1|\Psi_{j=1} \rangle = (|1\rangle + \sqrt{2}|2\rangle + |3\rangle)/2$ is lower in energy by $\Delta \propto t_{\textrm{inter}}$ from the non-bonding eigenstate, $^1|\Psi_{j=2} \rangle = (|1\rangle + |3\rangle)/\sqrt{2}$. This energy gap determines the time-period of the coherent triplet-pair population oscillations described in Section \ref{Se:5.1}.


\begin{figure}
\includegraphics[width=1.1\linewidth]{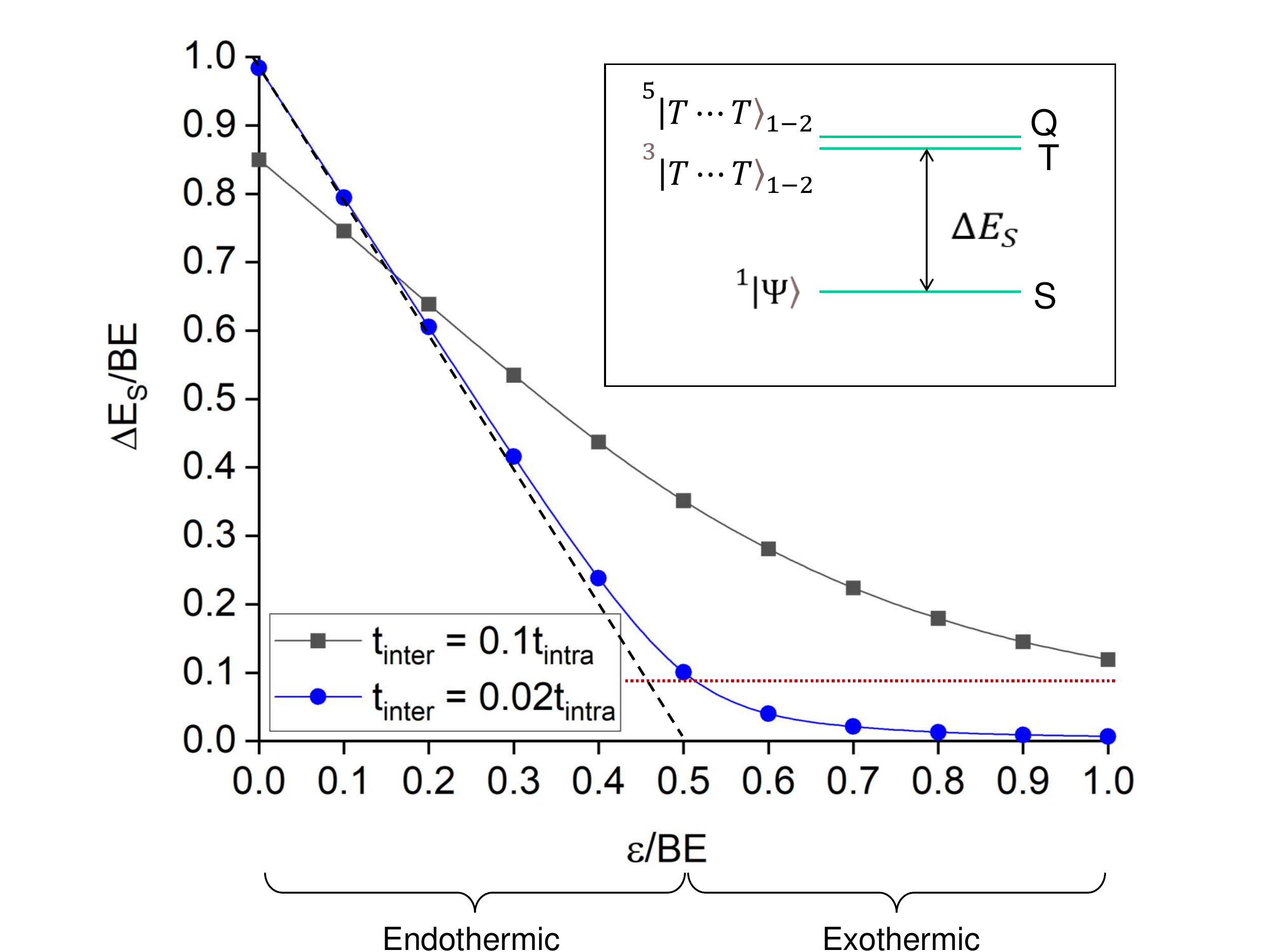}
\caption{The two-chain exchange energy, $\Delta E_S$, versus the offset-energy, $\varepsilon$, for weak and intermediate interchain coupling, \tinter.
$\Delta E_S$, illustrated in the inset, is the energy gap between the lowest energy singlet triplet-pair eigenstate (eqn (\ref{Eq:1140})) of the full two-chain Hamiltonian  (eqn (\ref{Eq:400})) and the (predominately) interchain triplet and quintet triplet-pair states, \TTTsep\ and \QTTsep.
The black dashed-line shows $\Delta E_S(t_{\textrm{inter}}=0) = (\textrm{BE}-\varepsilon/2)$, where BE (eqn (\ref{Eq:BE})) is the energy to dissociate an intrachain singlet triplet-pair  into two noninteracting,  interchain triplets  in the limit $t_{\textrm{inter}}=0$.
$\varepsilon < \textrm{BE}/2$ ($\varepsilon > \textrm{BE}/2$) corresponds to  endothermic (exothermic) singlet fission when \tinter = 0.
The red horizontal dashed-line represents the value of $k_BT$ at T = 300 K. \tintra= 0.88 eV.
}
\label{Fi:8}
\end{figure}

Next, let us consider the relative energies of the lowest  singlet, triplet and quintet triplet-pair eigenstates of the full two-chain Hamiltonian.
As discussed in Section \ref{Se:3.2}, when \tinter\ $=0$  and \eps\  $=0$ the singlet triplet-pair is bound on a single chain with a binding energy, BE, given by eqn (\ref{Eq:BE}). In contrast, because of weaker intrachain exchange interactions, the triplet and quintet triplet-pair eigenstates are the noninteracting, interchain  pairs,  \TTTsep\ and \QTTsep, respectively.
Thus, when \tinter\ $=0$ and \eps\ $=0$, the binding energy, BE, is equal to  the exchange energy, $\Delta E_S$, which separates the interacting  singlet triplet-pair from the noninteracting, interchain triplet and quintet triplet-pairs.
As the offset energy, \eps,  increases the interchain pair energies decrease. When \tinter\ $=0$, $\Delta E_S = (\textrm{BE}-\varepsilon/2)$, which  vanishes at the degeneracy point and in the exothermic regime. $\Delta E_S$ is illustrated schematically in the inset of Fig.\ \ref{Fi:8} and is shown as a function of \eps\ when \tinter\ $=0$ as the dashed line.

As we have seen, when the chains are coupled the intra and interchain singlet triplet-pairs hybridize to give the singlet eigenstate, eqn (\ref{Eq:1140}). In contrast, the triplet and quintet triplet-pair eigenstates remain predominately interchain in character. This means that the exchange energy, $\Delta E_S$, varies with \tinter, because the intrachain triplet-triplet attraction  causes an effective interchain triplet-triplet attraction in the singlet state. This is illustrated for two values of \tinter\ in Fig.\ \ref{Fi:8}.

Notice that since the triplet and quintet interchain triplet-pairs are noninteracting, $\Delta E_S$ is also the energy to dissociate the singlet triplet-pair eigenstate into a pair of noninteracting triplets on separate chains, defined by,
\begin{equation}\label{Eq:Es}
  \Delta E_S =   E(V=0, \varepsilon ,t_{\textrm{inter}})- E(V>V_c, \varepsilon ,t_{\textrm{inter}}),
\end{equation}
where $E$ is the lowest singlet energy of $\hat{H}$.
Thus, in the presence of spin-dephasing, when $  \Delta E_S \lesssim k_B T$ the \TTsep, \TTTsep\ and \QTTsep\ states mix to become uncorrelated, noninteracting single triplets on each chain.
Fig.\ \ref{Fi:8} shows the value of $k_BT$ when T = 300 K. For weak interchain coupling (i.e., \tinter = 0.02\tintra) $\Delta E_S$ is smaller than $k_BT$ in the exothermic regime.

\section{Theoretical and Computational Methodology}\label{Se:4}

\subsection{The Quantum Liouville Equation}\label{Se:4.1}

The nonequilibrium dynamics of the quantum system is fully described by its time-dependent density operator, whose time-evolution is determined by the quantum Liouville equation.
We compute the evolution of the density operator in the eigenstate basis of the two-chain Hamiltonian, i.e., $\rho_{ab} = \langle a | \hat{\rho} |b\rangle$.
In this work we adopt the frequently-used secular approximation\cite{Nitzan2006,Kuhn2011}, which explicitly decouples the evolution of the populations (i.e., the diagonal elements of the density matrix) from the coherences (i.e., the off-diagonal elements of the density matrix).

For spin-conserving processes arising from nonmagnetic system-bath interactions, the quantum Liouville equation for the populations $P_a \equiv \rho_{aa}$ is,
\begin{equation}\label{Eq:114}
  \frac{d P_{a}}{dt} = - \sum_{b \ne a}\left( k_{a  b} P_{a} - k_{b  a} P_{b} \right)
\end{equation}
while for the coherences it is,
\begin{equation}\label{Eq:115}
  \frac{d \rho_{ab}}{dt} = -i \omega_{ab} \rho_{ab} - 2\Gamma_{ab}(1-\delta_{ab})\rho_{ab}.
\end{equation}

Defining the Bohr frequencies as $\omega_{ab} = (E_a - E_b)/\hbar$ and taking $\omega_{ab} \geq 0$, the thermal rates are,\cite{Nitzan2006,Kuhn2011}
\begin{equation}\label{Eq:106}
  k_{a  b} = \left(\frac{2 \lambda}{\hbar} \right) J(\omega_{ab})(n(\omega_{ab}) + 1) C_{ab}
\end{equation}
and
\begin{equation}\label{Eq:107}
  k_{ b a} = \left(\frac{2 \lambda}{\hbar} \right) J(\omega_{ab})n(\omega_{ab}) C_{ab},
\end{equation}
where $n(\omega) = (\exp \beta\hbar\omega -1)^{-1}$ is the Bose distribution function
and   $J(\omega) = \omega \omega_0/(\omega^2+\omega_0^2)$ is the (dimensionless) Debye-spectral function.

We note that the thermal rates satisfy detailed balance, i.e., $ k_{ b a}/ k_{ ab} = \exp(-\beta\hbar\omega_{ab})$, ensuring that the steady-state eigenstate populations satisfy the Boltzmann distribution.
$\lambda$ is the bath reorganization energy while $C_{ab} = 2\sum_{ij} S_{ia}^2 S_{jb}^2$, where $\textbf{S}$ is the matrix whose columns are the eigenvectors of the two-chain Hamiltonian  represented in the `site' basis. $2\Gamma_{ab} = (\gamma_a + \gamma_b)$, where $\gamma_a = \sum_b  k_{a  b}$.

We also include magnetic system-bath interactions, induced for example by electron-nuclear magnetic dipole interactions. We will consider spin-dephasing. This  causes transverse relaxation and spin interconversion, i.e., total $S_z$ is conserved but total $S^2$ is not. We supplement the  quantum Liouville equation with a Lindblad dissipator\cite{Breuer2002,Marcus20}, where the Lindblad operator acting on each electron is $\hat{L} = \hat{S}_z/\hbar$. Using the $S_z=0$ triplet-pair bases states, we derive the additional population equations of motion as,
\begin{equation}\label{}
  \frac{d P_{a}^S}{dt} = - \frac{2 \gamma}{3}\sum_{b\in T}\left( k_{a  b} P_{a}^S - k_{b  a} P_{b}^T \right),
\end{equation}
\begin{eqnarray}\label{}
  \frac{d P_{a}^T}{dt} =&& - \frac{2 \gamma}{3}\sum_{b\in S}\left( k_{a  b} P_{a}^T - k_{b  a} P_{b}^S \right)
  \nonumber\\
  && - \frac{ \gamma}{3}\sum_{b\in Q}\left( k_{a  b} P_{a}^T - k_{b  a} P_{b}^Q \right)
\end{eqnarray}
and
\begin{equation}\label{}
  \frac{d P_{a}^Q}{dt} = - \frac{ \gamma}{3}\sum_{b\in T}\left( k_{a  b} P_{a}^Q - k_{b  a} P_{b}^T \right).
\end{equation}
Here, $P_a^S$, $P_a^T$ and $P_a^Q$ are the populations of eigenstate $|a\rangle$ in the singlet, triplet and quintet spin-sectors, respectively. In order to maintain detailed  balance, the rates $k_{ab}$ and $k_{ba}$ are given by eqn (\ref{Eq:106}) and eqn (\ref{Eq:107}). However, $\gamma$ is a multiplicative factor to account for the different strengths of the magnetic and nonmagnetic interactions. In this work we have taken $\gamma = 10^{-6}$.

The initial condition on the density operator is that  $\hat{\rho}(0) = |^1\Phi(0)\rangle\langle ^1\Phi(0)|$, where $|^1\Phi(0)\rangle$ is the lowest-energy singlet eigenstate of the single-chain Hamiltonian for chain $\times =1$, eqn (\ref{Eq:401}), i.e., it is the lowest energy member of the band of intrachain singlet triplet-pair states, \TT{1}, described in Section \ref{Se:3.1}.

\subsection{Observables}

In order to understand the singlet fission process, we calculate various observables. These are:

\begin{itemize}
\item{The Horodecki entanglement\cite{Horodecki2009,Marcus20} of the triplet-pair state, defined as,
\begin{equation}\label{}
  E_N = \log_2\| (\hat{\hat{1}}_A \otimes \hat{\hat{\tau}}_B ) \hat{\rho}_{AB} \|_{\textrm{tr}}.
\end{equation}
Here, we take subsystem A as one component of the triplet-pair, i.e.,  $A \equiv i$ and subsystem B as the other component, i.e.,  $B \equiv j$.
$\| \hat{O} \|_{\textrm{tr}}$ means the trace norm\footnote{The trace norm of $\hat{O}$ is the sum of square root of the eigenvalues of $\hat{O}\hat{O}^{\dagger}$.} of $\hat{O}$, while $\hat{\hat{\tau}}_B$ is the transposition superoperator for subsystem B.
The pair-entanglement vanishes at complete singlet fission.
}
\item{Site and eigenstate basis coherences, defined respectively as,
\begin{equation}\label{}
  C_{\textrm{site}} = 1 - \sum_{ij} |\tilde{\rho}_{ij}|
\end{equation}
and
\begin{equation}\label{}
  C_{\textrm{eigenstate}} = 1 - \sum_{ab} |{\rho}_{ab}|.
\end{equation}}
\item{Triplet-pair populations, $P_{\alpha}(t)$, in a projected subspace, $\alpha$. This is defined as,
\begin{equation}\label{}
  P_{\alpha}(t) = \textrm{Tr}\left\{ \hat{P}_{\alpha} \hat{{\rho}}(t) \right\},
\end{equation}
where the projection operator is
\begin{equation}\label{}
 \hat{P}_{\alpha} = \sum_{ij \in \alpha} |ij\rangle \langle ij|.
\end{equation}
Here, the projected sub-spaces of interest are the singlet, triplet and quintet pair-subspaces, as well as the additional projection of these subspaces onto a single or double chain. These projections allow us to identify the populations of the intrachain triplet-pair states, \TT{1}\ and \TT{2}, and the interchain triplet-pair states, $^{2S+1}|T \cdots T\rangle_{1-2}$ with a definite spin, $S$.}
\item{The expectation value of total spin,
\begin{eqnarray}\label{}
 \langle S^2 \rangle = && \textrm{Tr}\{  \hat{S}^2 \hat{\rho} \}
 \nonumber\\
 = && \hbar^2( 0\times P_S + 2 \times P_T + 6 \times P_Q).
\end{eqnarray}
At complete singlet fission, $P_S = P_T  = P_Q  = 1/3$ (for $S_z$ conserving spin-dephasing) and $\langle S^2 \rangle = 8\hbar^2/3$.
}
\end{itemize}

\subsection{Numerical Techniques}\label{Se:4.2}

The decoupling of populations and coherences via the secular approximation considerably reduces the complexity of solving the quantum Liouville equation.  This is a significant advantage, as even for this reduced basis model the Hilbert space of this problem is large.

The population equations of motion, eqn (\ref{Eq:114}), can be cast into the general form,
\begin{equation}\label{}
  \frac{d P_{a}}{dt} =  \sum_{\forall b} K_{ab} P_b,
\end{equation}
where $\textbf{K}$ is the matrix of the rate constants.
The solution from linear algebra is
\begin{equation}\label{}
  P_a(t) = \sum_{b c} S_{ab} \exp (\lambda_b t) S_{bc}^{-1} P_c(0),
\end{equation}
where $\textbf{S}$ is the matrix whose columns are the eigenvectors of $\textbf{K}$, $\{ \lambda \}$ are the corresponding eigenvalues and $P_c(0)$ is an initial condition.

Equation (\ref{Eq:115}) has the simple solution,
\begin{equation}\label{Eq:30}
  \rho_{ab}(t) =  \rho_{ab}(0)\exp\left ((-i \omega_{ab}  - 2\Gamma_{ab})t\right).
\end{equation}

\subsection{Parameters}

\begin{table}[h]
\small\centering
{\renewcommand{\arraystretch}{1.2}
\begin{tabular}{|p{4cm}|p{4cm}|}
\hline
Parameter &  Value \\
\hline

\tintra, eqn (\ref{Eq:401}) & 0.88 eV \\
No.\ of C-atoms per chain, $N$ & 20 \\
BE$_{\textrm{single}}$ &    0.5637\tintra = 0.50 eV \\
BE &    0.3272\tintra = 0.29 eV \\
Singlet triplet-pair attraction, $V_S=J/2$, eqn (\ref{Eq:401}) & 2.8\tintra = 2.5 eV\\
Reorganization energy, $\lambda$ & 0.05 eV \\
Temperature, $T$ &  300 K \\
\hline
\end{tabular}}
\caption{Values of parameters used in this paper. (See also Table II.)}
\label{Ta:2}
\end{table}

The intrachain triplet hopping matrix element, \tintra, can be estimated via the Pariser-Parr-Pople (or extended Hubbard) model of $\pi$-conjugated systems\cite{Book}. Intrachain triplet hopping between ethylene dimers is a second-order process, occuring via a virtual charge-transfer state. According to ref\cite{Barford2022c},
the superexchange transfer integral is
 $ t_{\textrm{intra}} = - \alpha^2 \beta_s^2/\Delta_{\textrm{CT}}$,
where $\Delta_{\textrm{CT}}$ is the energy gap between the local and virtual charge-transfer triplet states, $\beta_s $ is the one-electron transfer integral across a single bond, and $\alpha^2$  is the probability that the neighboring singlet dimer is in the covalent state (shown as a lozenge in Fig.\ \ref{Fi:18}). Using Pariser-Parr-Pople parameters, we estimate \tintra\ to be ca.\ 0.88 eV\cite{Barford2022c}.

Interchain triplet hopping  is also a two-electron virtual process. The values of the one-electron interchain transfer integrals depend sensitively on interchain separations and conformation.
According to refs\cite{Reichman2013b,Ghosh2022} typical values for conjugated molecule dimers are 0.2 - 0.4 eV, which are 3 - 6 times smaller than the corresponding intrachain interdimer one-electron transfer integrals. This implies that the interchain  superexchange is 9 - 36 times smaller than the intrachain  superexchange, and thus we take \tinter\ as a variable, satisfying 0.1\tintra\ $\le$ \tinter $\le$ 0.01\tintra.

As shown in Appendix A, according to the valence-bond model, the singlet triplet-pair exchange interaction, $V_S = 2J =$ 2\tintra. As also discussed in Section \ref{Se:3.1} and shown by Fig.\ \ref{Fi:16}, this would imply that the triplet-pair is not bound in polyenes, an observation at variance with experiment\cite{Clark2023} and rigorous DMRG calculations\cite{Valentine20}.
This discrepancy is a consequence of using the reduced valence-bond basis, which has no ionic contributions\cite{Barford2022c}. However, as it is theoretically and computationally expedient to use this minimal basis, we account for the discrepancy by treating $V_S$ as an adjustable parameter.
We thus set $V_S = 2.8$\tintra, which (as shown in Fig.\ \ref{Fi:16}) reproduces the binding energy derived from the  enlarged valence-bond-exciton basis\cite{Barford2022c}. We set the reorganization energy to a value typical of conjugated molecular systems\cite{Reichman2013b}, namely $\lambda = 0.05$ eV, and there are 20 C-atoms per chain.

For convenience, the adjustable and derived parameters are listed in Table I.


\subsection{Glossary of Terms}

For convenience,  Table II contains a list of the terms and definitions used in this paper. In particular, it defines the different types of triplet-pair states that are the subject of the following section.

\begin{table}[h]
\small\centering
{\renewcommand{\arraystretch}{1.2}
\begin{tabular}{|p{2.5cm}|p{5.5cm}|}
\hline
Term &  Definition \\
\hline
\tintra & The intrachain triplet hopping matrix element. \\
\tinter & The interchain triplet hopping matrix element. \\
$J = V_S/2$ & The triplet-triplet exchange interation. \\
\eps & The interchain triplet  off-set energy.  2\eps $<$ BE (2\eps $>$ BE) implies potential endothermic (exothermic) singlet fission in the limit that \tinter =0. \\
\hline
$^1|TT\rangle_{\times}$ & A bound, \emph{intra}chain singlet triplet-pair localized on chain $\times$.
$^1|TT\rangle_{}$ is an eigenstate of the single chain Hamiltonian $\hat{H}_{\textrm{single}}$, eqn (\ref{Eq:401}).\\
$^{2S+1}|T \cdots T\rangle_{1-2}$ & A noninteracting, \emph{inter}chain triplet-pair with spin $S$.
$^{2S+1}|T \cdots T\rangle_{1-2}$  is an eigenstate of the double chain Hamiltonian $\hat{H}_{\textrm{double}}$, eqn (\ref{Eq:402}).\\
$^1|\Psi\rangle$         & Singlet eigenstate of the full two-chain Hamiltonian (eqn (\ref{Eq:400})), defined in eqn (\ref{Eq:1140}).\\
$|T\rangle_{\times}$ & A single triplet localized on chain $\times$. \\
\hline
BE$_{\textrm{single}}$ & The energy to dissociate the lowest energy \TT\ state into two non-interacting \emph{intra}chain triplets. It is defined by eqn (\ref{Eq:B7}) and illustrated in Fig.\ \ref{Fi:14}.
Notice that since triplets in the quintet pair state are unbound, BE$_{\textrm{single}}$ is also the intrachain singlet-quintet triplet-pair exchange energy. \\
BE & The energy to dissociate the lowest energy \TT\ state into two noninteracting, \emph{inter}chain triplets in the limit that \tinter = \eps = 0. It is defined by eqn (\ref{Eq:BE}). \\
$\Delta E_S$ &
The exchange energy between the lowest energy singlet triplet-pair eigenstate (eqn (\ref{Eq:1140})) of the full two-chain Hamiltonian (eqn (\ref{Eq:400}))  and the noninteracting, interchain triplet and quintet states, \TTTsep\ and \QTTsep.
$\Delta E_S$ is also the  energy to dissociate the lowest energy singlet triplet-pair eigenstate  into two noninteracting interchain triplets for general \tinter\ and \eps.
(Note that $\Delta E_S$(\eps=0, \tinter=)  $\equiv$  BE.)
It is defined by eqn (\ref{Eq:Es}) and shown in Fig.\ \ref{Fi:8}.
\\
\hline
\end{tabular}}
\caption{A glossary of terms and definitions used in this paper.}
\label{Ta:1}
\end{table}


\section{Results and Discussion}\label{Se:5}

We now turn to discuss our results, starting with a discussion of the dynamics of the triplet-pairs, before describing how the model parameters determine their equilibrium populations. We summarize our findings in Section \ref{Se:5.3}.

\subsection{Dynamics}\label{Se:5.1}

All the dynamical simulations were  computed  at the degeneracy point, i.e., $\varepsilon = \textrm{BE}/2$, with $t_{\textrm{inter}} = 0.02 t_{\textrm{intra}}$ and $t_{\textrm{intra}} = 0.88$ eV, and at a temperature $T = 300$ K.

\subsubsection{Unitary Evolution of the Closed Quantum System}\label{Se:5.1.1}

\begin{figure}[h]
\includegraphics[width=1.0\linewidth]{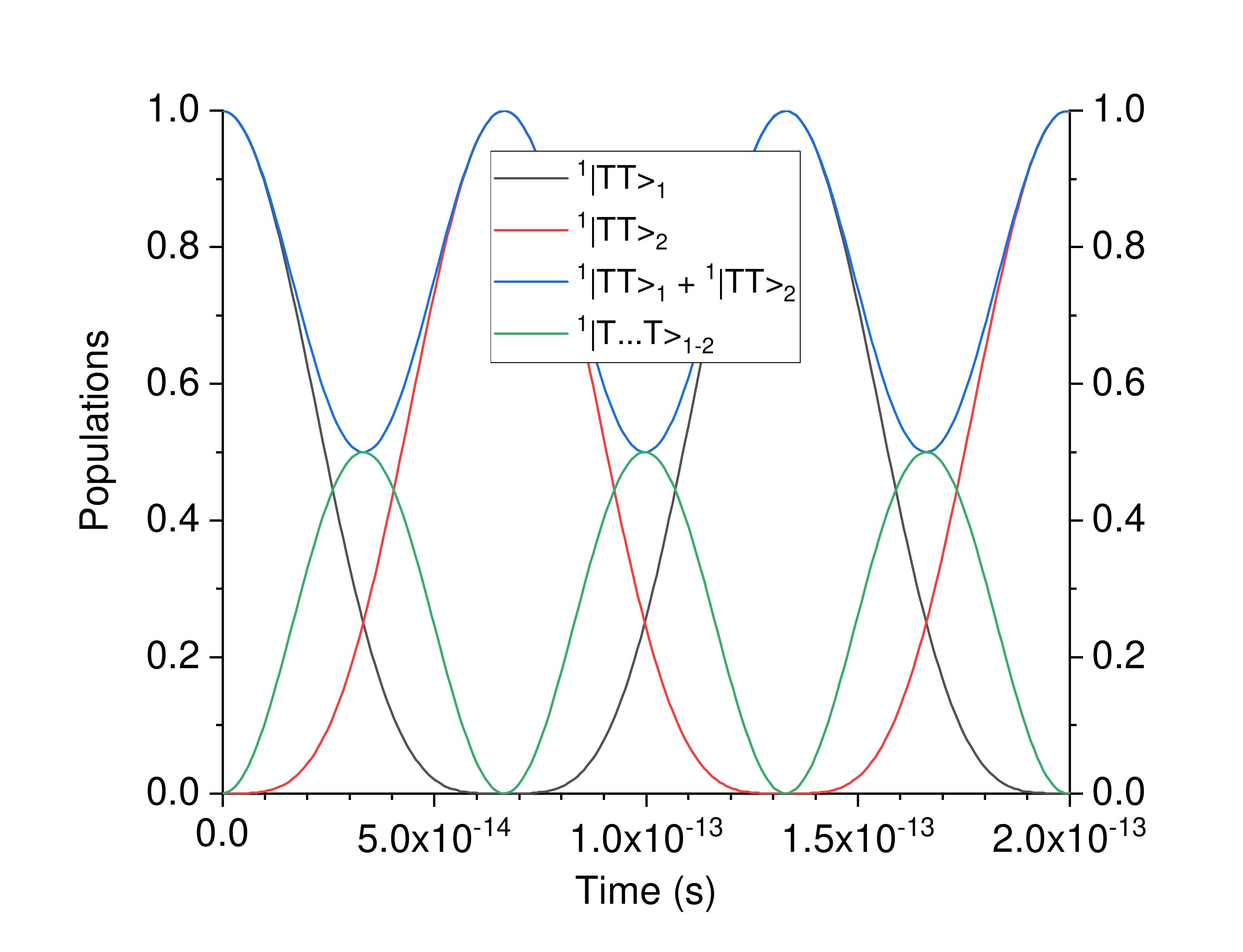}
\caption{The coherent oscillations of the triplet-pair populations in the absence of dephasing. Here, the time-period of the return probability (i.e., P(\TT{1})) is $\tau \propto 1/t_{\textrm{inter}}$. (See also Fig.\ \ref{Fi:2}.)}
\label{Fi:1}
\end{figure}

We begin  our discussion of the dynamics by investigating the triplet-pair populations in the unitary limit, i.e., when the evolution is only determined  by the time-dependent Schr\"odinger equation for a closed system. Fig.\ \ref{Fi:1} illustrates the coherent oscillations of the triplet-pair populations at the degeneracy point, $\varepsilon = \textrm{BE}/2$. The initial  \TT{1} population on chain 1  transfers  to chain 2 via the interchain \TTsep\ state.
At  33 fs P(\TTsep) = 0.5, while P(\TT{1}) = P(\TT{2}) = 0.25. At 66 fs the triplet-pair population has entirely transferred to chain 2. The return probability period, $\tau$, (i.e., the period for which P(\TT{1}) = 1) is 132 fs. This period is determined by the energy gap between the bonding and nonbonding two-chain singlet triplet-pair eigenstates, discussed in Section \ref{Se:3.3}, i.e., by $\Delta \propto t_{\textrm{inter}}$, and thus $\tau \propto 1/t_{\textrm{inter}}$.

\begin{figure}[h]
\includegraphics[width=1.0\linewidth]{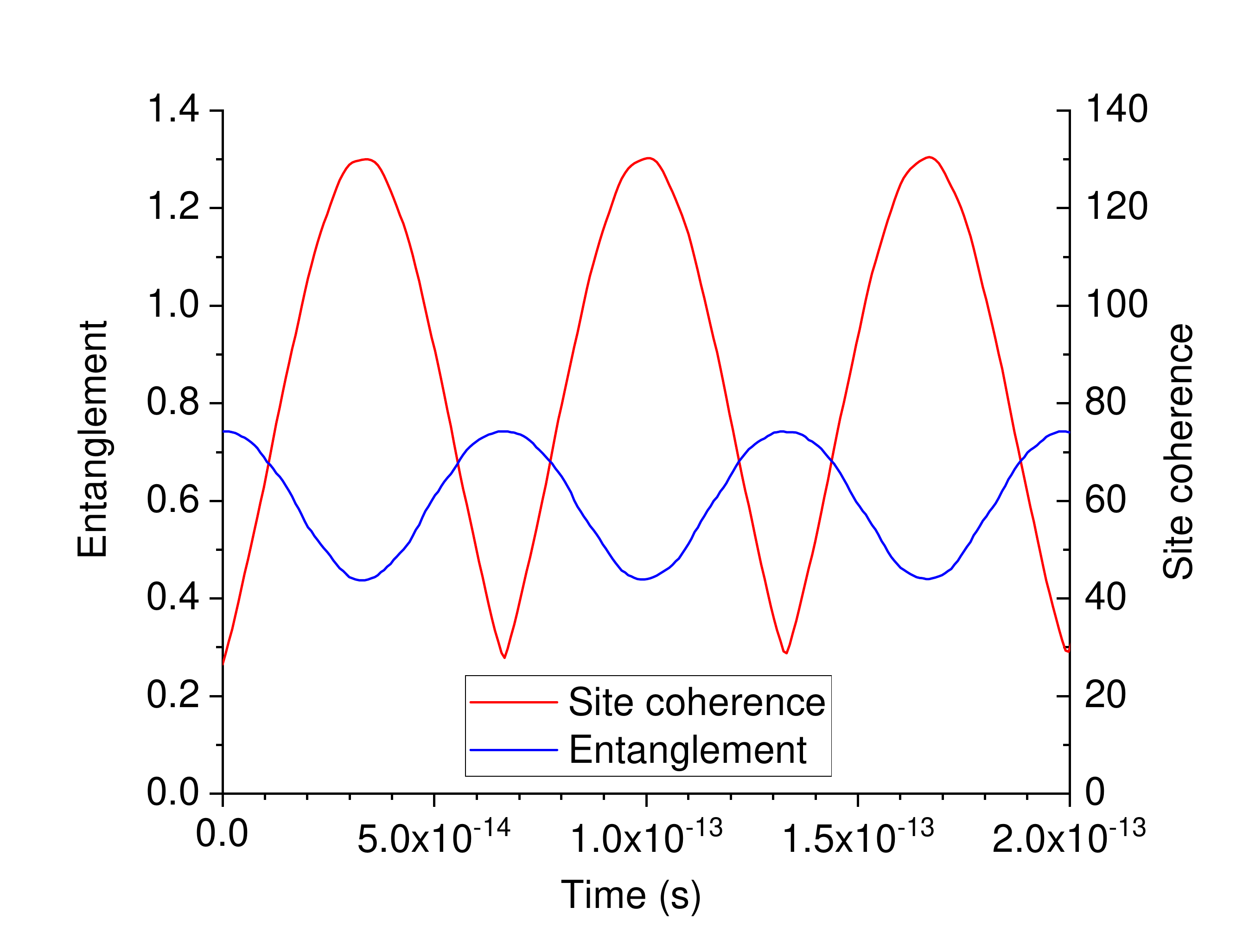}
\caption{The oscillations of the site-coherence and the Horodecki entanglement in the absence of dephasing. These are maximized and minimized, respectively, when the inter-chain population, \TTsep, is maximized. (See also Fig.\ \ref{Fi:1}.)}
\label{Fi:2}
\end{figure}

Associated with the coherent population dynamics are oscillations in the site-coherence and the Horodecki entanglement, as illustrated in Fig.\ \ref{Fi:2}. As expected, the site coherence is maximized when  P(\TTsep) is maximized and P(\TT{}) are minimized. Conversely, the entanglement is minimized at these times, indicating that the intrachain triplet pairs, \TT{}, are more entangled than the interchain pairs, \TTsep. Indeed, noting that the pair-entanglement is 0.74 at $t=0$ and 0.44 (and not 0.37) at $t=\tau/4$, we deduce that the Horodecki entanglement of \TT{}\ and \TTsep\ are 0.74 and 0.14, respectively.

\subsubsection{The Role of Dissipation}\label{Se:5.1.2}

\begin{figure}[h]
\includegraphics[width=1.1\linewidth]{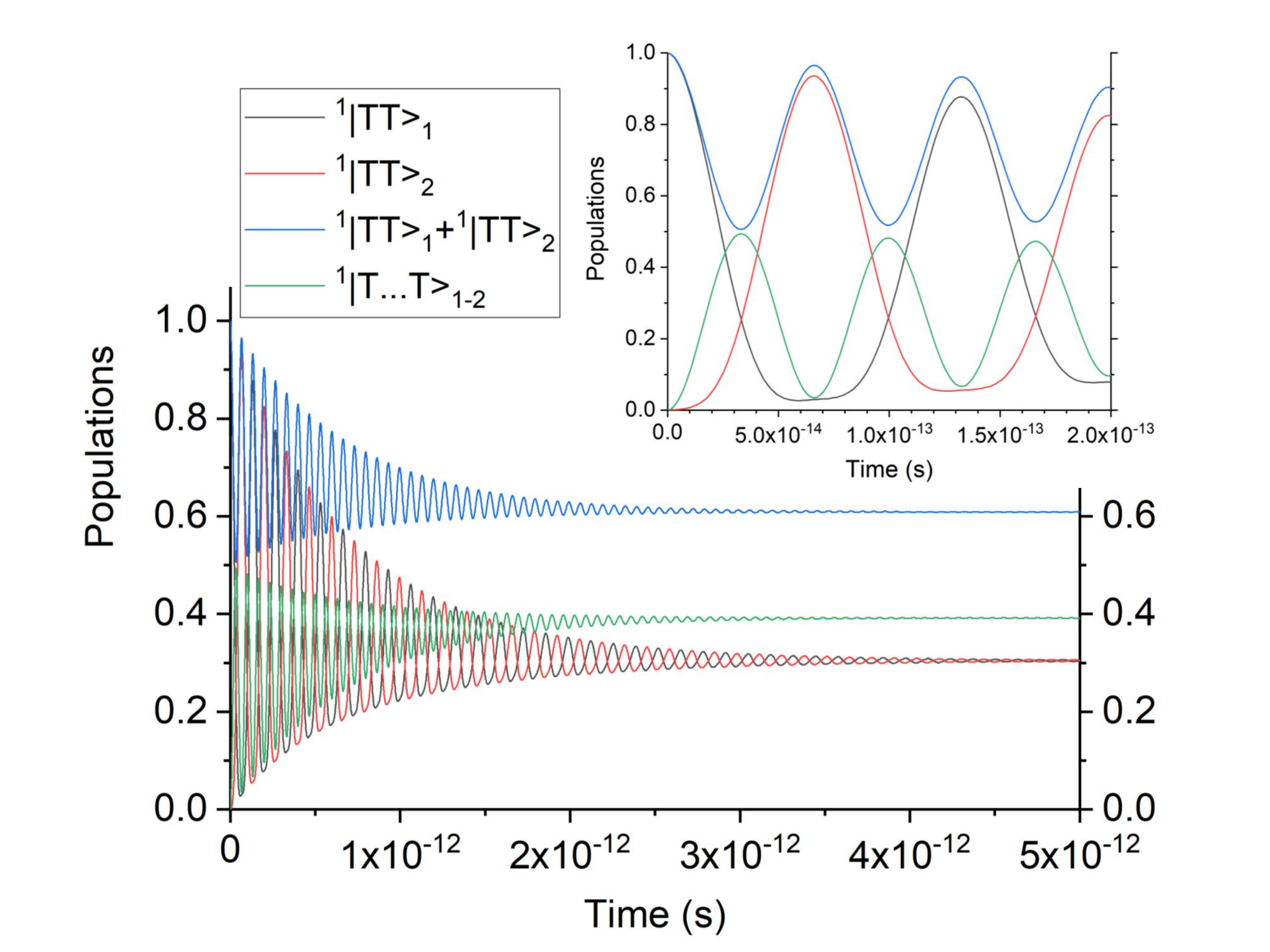}
\caption{The populations as a function of time without spin-dephasing of the intrachain singlet triplet-pair states, \TT{1} and \TT{2}, and the interchain singlet triplet-pair state, \TTsep. The populations have thermalized within 5 ps. The inset shows the ultrafast population dynamics.}
\label{Fi:3}
\end{figure}

We now turn to the role of thermal dissipation in the absence of spin-dephasing. Again, Fig.\ \ref{Fi:3} shows the triplet-pair populations as a function of time at the degeneracy point. Comparing the inset of  Fig.\ \ref{Fi:3} with  Fig.\ \ref{Fi:1}, we see that after each return period, $\tau$, the \TTsep\ population  accumulates while the net \TT{} population decreases. The oscillation of these populations are damped,   reaching equilibrated values of P(\TT{}) $\sim 0.6$ and P(\TTsep) $\sim 0.4$  within 5 ps.

\begin{figure}[h]
\includegraphics[width=1.0\linewidth]{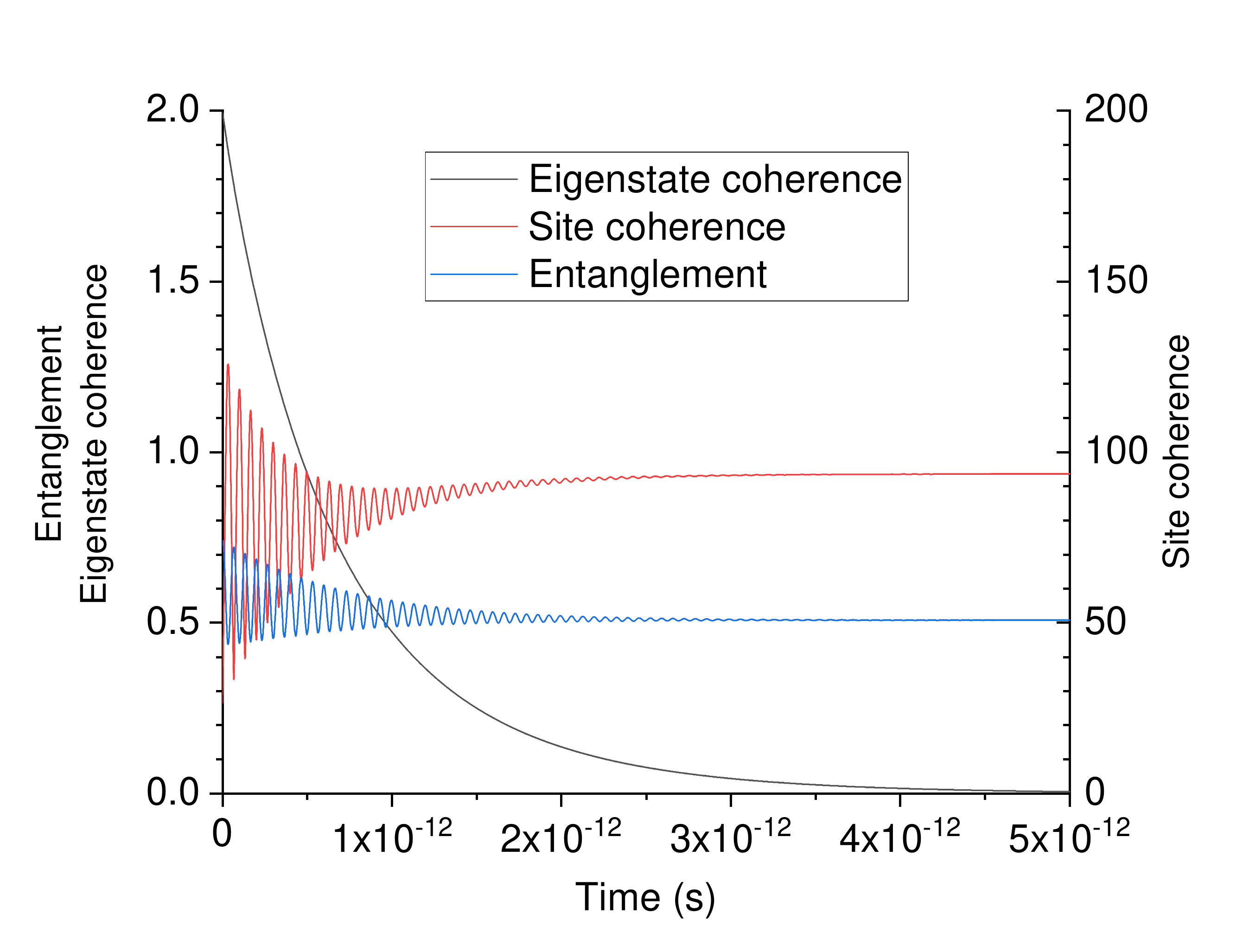}
\caption{The site and eigenstate coherences, and the entanglement as a function of time.  The decay time of the eigenstate coherence is $\sim 1$ ps.}
\label{Fi:4}
\end{figure}

The coherences and entanglement are displayed in Fig.\ \ref{Fi:4}. As expected from eqn (\ref{Eq:30}), the eigenstate coherence decreases as a sum of exponentials; a single-exponential fit gives a decay time $\sim 1$ ps. The eigenstate coherences are negligible after 5 ps, meaning that population of the eigenstates of the full two-chain Hamiltonian have achieved their (classical) Boltzmann values.

The oscillations of the site-coherence and entanglement are damped, again reaching steady-state values within 5 ps. Time-averaging these oscillations over one period, the time-averaged entanglement decreases monotonically from an initial value of 0.74  to a final value of 0.51. This indicates that although population has transferred from \TT{} to \TTsep, the \TTsep\ state is still an entangled pair. Conversely, the time-averaged site-coherence increases monotonically, from 76 to 94, indicating that this is not a good measure of the `quantumness' of the entangled pair.

Some authors define the creation of a population of \TTsep\ states as singlet fission\cite{Zhu2019}, partly because for some spectroscopies spin-correlated, but noninteracting triplet-pairs are indistinguishable from individual triplets\cite{Musser19,Zhu2019}. However, from a theoretical perspective, this definition is unsatisfactory as in general the \TTsep\ component is part of the correlated singlet triplet-pair eigenstate, $^1|\Psi\rangle$ (eqn (\ref{Eq:1140})). As described in Section \ref{Se:3.3}, it costs an energy $\Delta E_S$ to dissociate (i.e., split or divide) this state into separate, uncorrelated triplets. We return to our definition of singlet fission in Section \ref{Se:5.3}.

\subsubsection{The Role of Spin Dephasing}\label{Se:5.1.3}

\begin{figure}[h]
\includegraphics[width=1.1\linewidth]{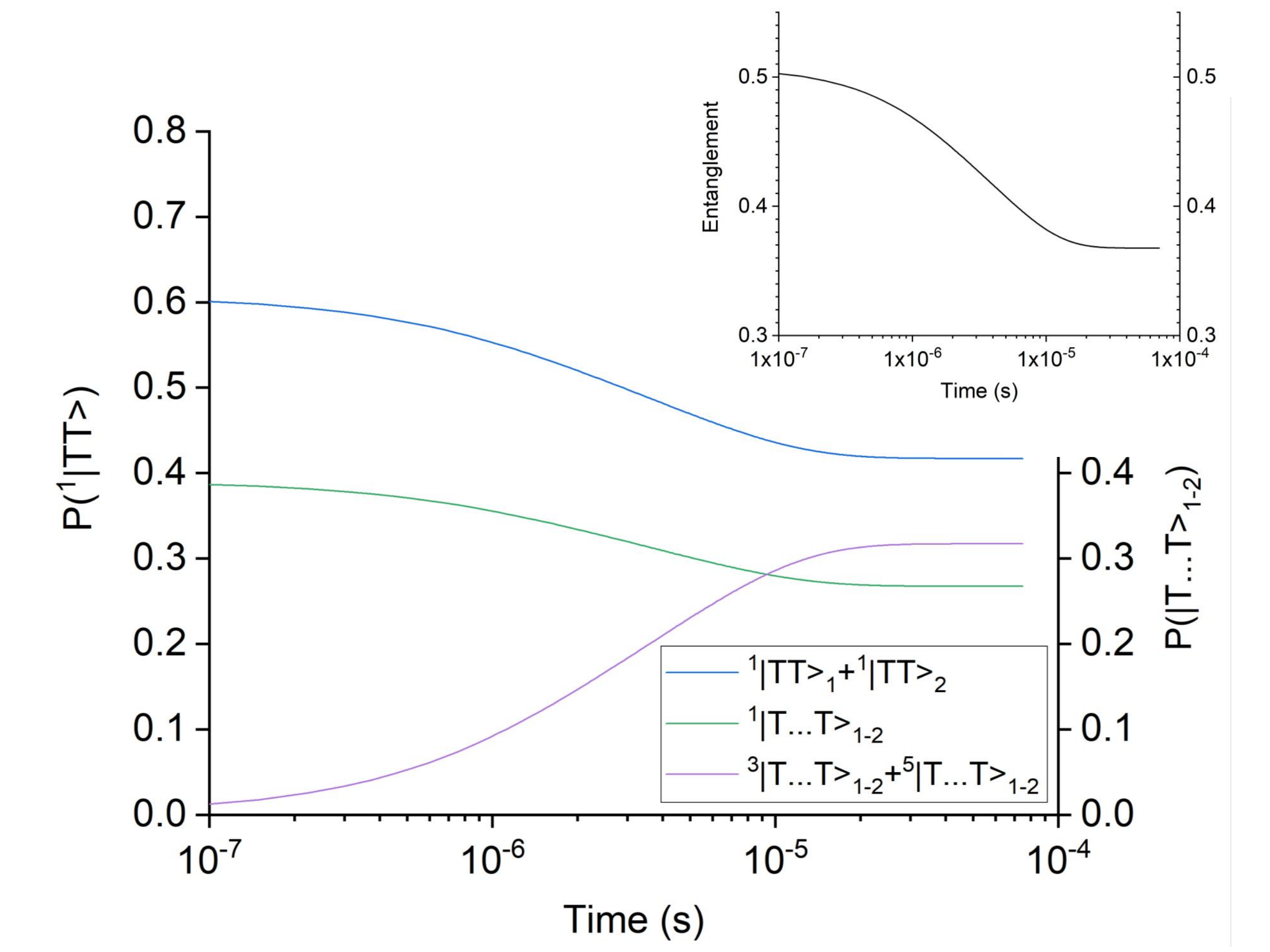}
\caption{The populations as a function of time with spin dephasing of the intrachain singlet triplet-pair states, \TT,  and the interchain singlet triplet-pair state, \TTsep. Also shown is the joint population of  the  interchain triplet and quintet triplet-pair states, \TTTsep\ and \QTTsep.
The spin dephasing rate is $10^{-6}$ times smaller than the Redfield thermalization rate.
These results show that after thermalization of the \TT{\times} and \TTsep\  populations in less than 5 ps (as illustrated Fig.\ \ref{Fi:3}) slower spin-dephasing converts some of the \TT{\times} and \TTsep\ populations into  \TTTsep\ and \QTTsep\ (or equivalently, \T{1} and \T{2}). Associated with this interconversion of the spin-correlated triplet-pairs to uncorrelated triplets is a reduction of the pair-entanglement, as illustrated by the inset. }
\label{Fi:5}
\end{figure}

Finally in our description of the dynamics, we turn to role of spin-dephasing. To distinguish the kinetics of this process from the spin-conserving thermalization described in Section \ref{Se:5.1.2}, we take a spin-dephasing rate $10^{-6}$ times smaller, implying a spin-relaxation time of ca.\ $1\ \mu$s. Spin-dephasing causes population transfer from the interchain singlet triplet-pairs to the interchain triplet and quintet triplet-pairs, i.e., to the $S_z = 0$ components of \TTTsep\ and \QTTsep\ from \TTsep. However, as \TTTsep\ and \QTTsep\ are quasidegenerate they are mixed by spin-dephasing, implying that they are equivalent to uncorrelated, single triplets on each chain.

The triplet populations are illustrated in Fig.\ \ref{Fi:5}. These results show that after the thermalization of the \TT{} and \TTsep\  populations in less than 5 ps (at $\sim 0.60$ and $\sim 0.40$, respectively, as illustrated Fig.\ \ref{Fi:3}) slower spin-dephasing converts some of the \TT{} and \TTsep\ population to \TTTsep\ and \QTTsep. The equilibrated \TT{} and \TTsep\  populations are now $ 0.42$ and $ 0.27$, respectively, while the  \TTTsep\ and \QTTsep\ populations are equal and sum to $0.31$. We remark that for these parameters the two-chain exchange energy, $\Delta E_S$, is (just) greater than $k_B T$. Thus, the lowest energy singlet triplet-pair eigenstate (given by eqn (\ref{Eq:1140})) does not readily interconvert with \TTTsep\ and \QTTsep.

Associated with the interconversion of the spin-correlated triplet-pairs to uncorrelated triplets is a reduction of the pair-entanglement, as illustrated by the inset of Fig.\ \ref{Fi:5}, decreasing from 0.50 at 0.1 $\mu$s to  a final value of 0.37.

\subsection{Equilibrium Properties}\label{Se:5.2}

\begin{figure}[h]
\includegraphics[width=1.0\linewidth]{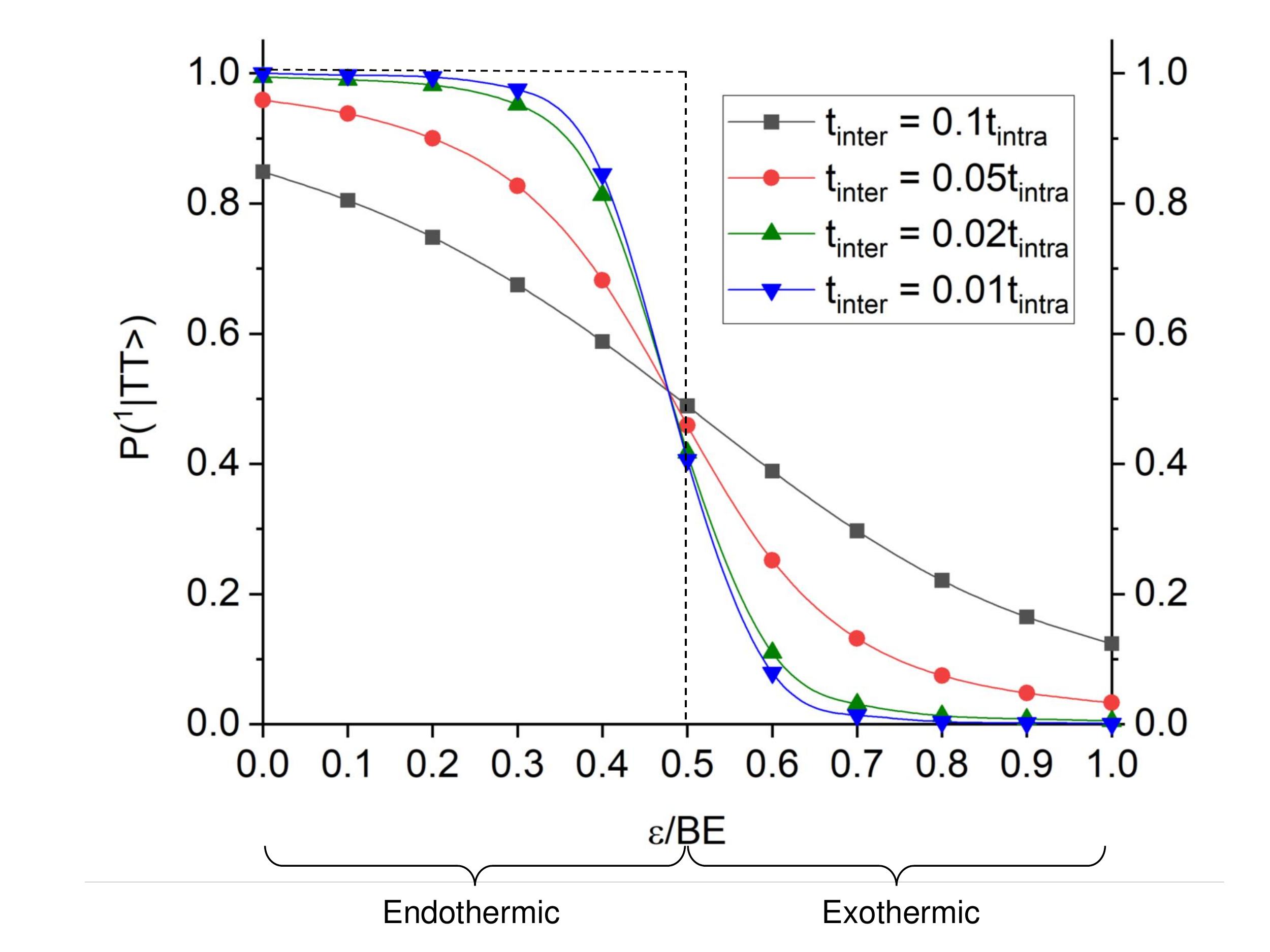}
\caption{The equilibrated intrachain singlet triplet-pair, \TT, population as a function of the offset-potential, $\varepsilon$, for different \tinter. The stepfunction, represented by the dash-lines, is the result for \tinter = 0 showing that an energy-level crossing occurs at $\varepsilon = \textrm{BE}/2$. This becomes an avoided crossing for \tinter $ \ne 0$. $\varepsilon < \textrm{BE}/2$ ($\varepsilon > \textrm{BE}/2$) corresponds to possible endothermic (exothermic) singlet fission, as explained in Section \ref{Se:5}.}
\label{Fi:6}
\end{figure}

The time taken for the spin populations to thermally equilibrate evidently depends on the rates of dissipation and dephasing. In this section we assume that equilibration has occured and we now discuss the equilibrium populations in order to understand the fate of the initial entangled pair.

Fig.\ \ref{Fi:6} illustrates the intrachain singlet triplet-pair, \TT{}, populations as a function of the offset-potential, $\varepsilon$, for different \tinter. For all our results the temperature T = 300 K. In the \tinter$ = 0$ limit there is an energy-level crossing at $\varepsilon = \textrm{BE}/2$ between a singlet eigenstate with triplet-pairs on either chain to one with the triplet-pair separated on both chains. This is illustrated by the dashed step-function in Fig.\ \ref{Fi:6}. At \tinter $ = 0$, for $\varepsilon < \textrm{BE}/2$ the triplet-pair is on either chain 1 or 2 and P(\TT{})  = 1. Conversely, for $\varepsilon > \textrm{BE}/2$ the triplet-pair is separated on both chains and P(\TT{})  = 0. Furthermore, as shown in Fig.\ \ref{Fi:8}, for $\varepsilon \ge \textrm{BE}/2$ the exchange energy, $\Delta E_S$, vanishes when \tinter $=0$. Therefore, if spin-dephasing is present the singlet, triplet and quintet interchain triplet-pairs mix, implying complete singlet fission, i.e., P(\T{})  = 2.
For \tinter$ \ne 0$, on the other hand, there is an avoided crossing at $\varepsilon = \textrm{BE}/2$ and the stepfunction is smeared out with increasing \tinter, as illustrated in Fig.\ \ref{Fi:6}.

\begin{figure}[h]
\includegraphics[width=1.0\linewidth]{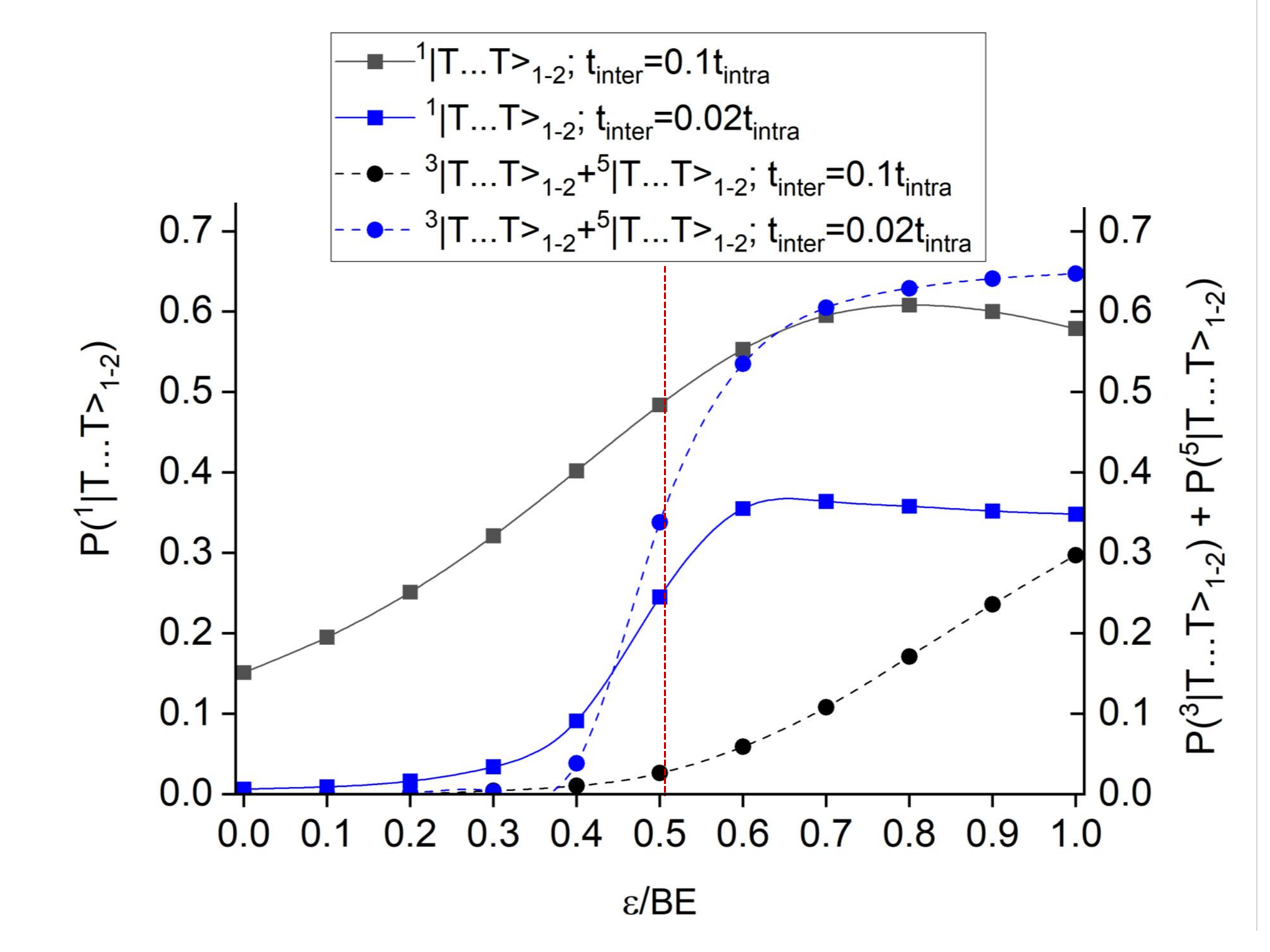}
\caption{The equilibrated populations of the interchain singlet triplet-pair state, \TTsep, and the sum of the populations of the interchain triplet and quintet triplet-pair states, \TTTsep\ and \QTTsep, as a function of the offset-potential, $\varepsilon$.
The red vertical dashed-line represents the value of $\varepsilon$ at which $\Delta E_S = k_BT$ for \tinter  = 0.02\tintra. It therefore indicates the crossover at which  \TTsep\ mixes with  \TTTsep\ and \QTTsep, which together populate the uncorrelated \T{1} and \T{2} states.}
\label{Fi:7}
\end{figure}

The interchain triplet-pair population may also be inferred from Fig.\ \ref{Fi:6}, as it is given by ($1-$ P(\TT{})). Depending on the parameter regimes and temperature, this population becomes a mixture of interchain singlet triplet-pairs, \TTsep, and uncorrelated triplets, \T{}.
We can understand this more fully via Fig.\ \ref{Fi:8} and Fig.\ \ref{Fi:7}.
The solid curves with square symbols in Fig.\ \ref{Fi:7} indicate the \TTsep\ populations as a function of the offset-potential for weak (i.e., \tinter =0.02 \tintra) and intermediate (i.e., \tinter\ = 0.1 \tintra) interchain coupling.
As shown in Fig.\ \ref{Fi:8}, for intermediate interchain coupling the exchange energy, $\Delta E_S$, is larger than $k_BT$. Nonetheless, for $\varepsilon > \textrm{BE}/2$ thermal excitation into the triplet and quintet sectors is possible, as indicated by the black dashed curve with circle symbols in Fig.\ \ref{Fi:7}. As these states are quasidegenerate, they mix forming single triplets on separate chains.

Conversely, as also shown in Fig.\ \ref{Fi:8}, for weak interchain coupling the exchange energy, $\Delta E_S$, is smaller than $k_B T$ when $\varepsilon > \textrm{BE}/2$. In this regime the singlet, triplet and quintet interchain pairs mix to form uncorrelated triplets on each chain. This represents complete singlet-fission of the initial \TT{1} state, and is indicated in Fig.\ \ref{Fi:7} by P(\TTsep) $\simeq$ P(\TTTsep) $\simeq$ P(\QTTsep) $\simeq 1/3$ at $\varepsilon/\textrm{BE} \gtrsim 0.6$.

\subsection{Singlet Fission: The Fate of the Entangled Pair}\label{Se:5.3}

\begin{figure}[h]
\includegraphics[width=1.0\linewidth]{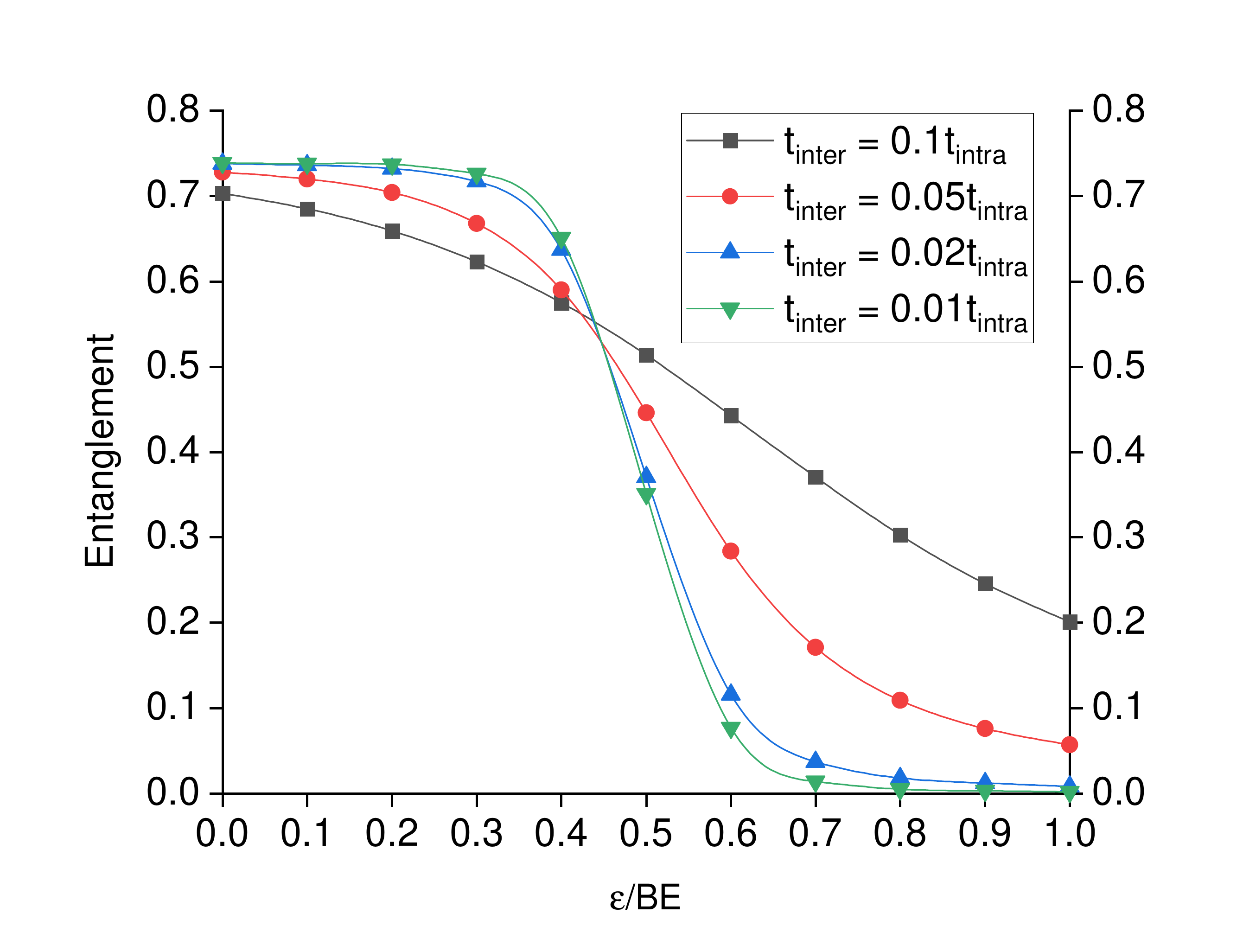}
\caption{The equilibrated triplet-pair entanglement as a function of the offset-potential, $\varepsilon$, for different \tinter. The entanglement vanishes as the single-triplet population on separate chains $\rightarrow 2$.}
\label{Fi:10}
\end{figure}

As described in previous papers\cite{Valentine20,Barford2022c}, the singlet triplet-pair  of carotenoids is one component of the `dark' state, which is born within tens of fs following photoexcitation of the `bright' state. (The other component of the `dark' state is a charge-transfer exciton. Although this has the important role of causing triplet-triplet attraction, it is not relevant for the rest of the present discussion.)

The initial singlet triplet-pair wavefunction is given by eqn (\ref{Eq:01}) and eqn (\ref{Eq:02}), indicating that it is an entangled pair of triplets. This paper has described the subsequent fate of this pair, showing that it depends intricately on a range of microscopic parameters and  temperature. The two key Hamiltonian parameters are the interchain coupling, \tinter, and the potential energy offset, \eps. A third parameter, the intrachain coupling \tintra, sets the overall energy scale, and hence determines the timescales of the dynamics. The detailed kinetics of singlet fission depend on the various dissipation and spin-dephasing parameters, while the final equilibrium populations depend on the Hamiltonian parameters and temperature.

The entangled pair, itself, exists as two types. The first type is the strongly-bound intrachain pair, \TT{}. As shown in Fig.\ \ref{Fi:1}, this pair oscillates between both chains with a time period determined at the degeneracy point by \tinter. During this transfer of population, the second type of entangled pair is created, namely the noninteracting, interchain pair, \TTsep. As shown by Fig.\ \ref{Fi:2}, \TTsep\ is less entangled than \TT.

In the absence of external dissipation, this nonstationary state dynamics is fully described by Fig.\ \ref{Fi:1}. As shown in Fig.\ \ref{Fi:3}, however, spin-conserving thermalization causes a time-averaged population to transfer from \TT{} to \TTsep. In the absence of spin-dephasing, \TTsep\ retains its spin coherence (its pair entanglement is 0.14 versus 0.74 for \TT{}) and thus true singlet fission into separate, uncorrelated triplets has not occurred.

The final step of singlet fission  occurs during spin-dephasing. As shown in Fig.\ \ref{Fi:4}, spin-dephasing causes population to transfer from \TTsep\ to interchain triplet and quintet triplet-pair states, i.e., \TTTsep\ and \QTTsep. Since the triplet and quintet triplet-pair states are quasidegenerate, they are mixed by spin-dephasing to form single, uncorrelated triplets on each chain, i.e., \T{1}\ and \T{2}. Assuming that spin-dephasing is slower than thermalization of the \TT{} and \TTsep\ states, the kinetic scheme is therefore\cite{Scholes2015}:
\begin{equation}\label{}
  ^1|TT\rangle_{} \rightarrow  {^1}|T \cdots T\rangle_{1-2} \rightarrow |T\rangle_{1} + |T\rangle_{2}.
\end{equation}

As already mentioned, and as illustrated in Fig.\ \ref{Fi:6} and Fig.\ \ref{Fi:7}, the equilibrated populations depend  on the Hamiltonian parameters and temperature. For weak interchain coupling (i.e., \tinter\ $\lesssim$ 0.02\tintra) there is a relatively sharp singlet-fission transition at the degeneracy point, where for \eps\ $<$ BE/2 (i.e., the endothermic regime) most of the triplet-pair population is \TT{}, whereas for \eps\ $>$ BE/2 (i.e., the exothermic regime) the triplets are an equal population of \TTsep, \TTTsep\ and \QTTsep\ (or equivalently, \T{1} and \T{2}). As shown in Fig.\ \ref{Fi:10}, this transition is accompanied by a sharp reduction in the pair-entanglement.

Conversely, for stronger interchain coupling (i.e., \tinter\ $\approx$ 0.1\tintra) there is a smoother crossover from predominately \TT{} population to predominately both \TT{} and \TTsep\ population to predominately both \TTsep\ and \T{} population as \eps\ is increased from the endothermic to exothermic regimes. In this case, the pair-entanglement slowly decreases as \eps\ increases.

As this discussion indicates, the pair-entanglement, illustrated in Fig.\ \ref{Fi:10} for a range of parameters, provides a useful measure of the `quantumness' of the triplet-pair, being maximized for the intrachain entangled-pair given by eqn (\ref{Eq:01}) and vanishing for separate, uncorrelated triplets.

In conclusion, complete singlet fission of \TT{}\ occurs when the populations of \TTsep, \TTTsep\ and \QTTsep\ are fully thermalized and equal 1/3 each (for the case of pure spin-dephasing and thus conserved $S_z$). (Equivalently, using the uncoupled representation, the populations of $|1,i_1\rangle|-1,j_2\rangle$, $|0,i_1\rangle|0,j_2\rangle$ and $|-1,i_1\rangle|1,j_2\rangle$  equals 1/3 each\cite{Scholes2015}.) Then, the populations of \T{1} and \T{2} equals 1 each, $\langle S^2 \rangle = 8\hbar^2/3$ and the Horodecki pair-entanglement vanishes. This happens, for example, for \tinter\ = 0.02\tintra\ deep in the exothermic regime, as shown in Fig.\ \ref{Fi:7} and Fig.\ \ref{Fi:10} at  \eps\ = BE.


\section{Concluding Remarks}\label{Se:4}

As described in the Introduction, singlet fission in carotenoids is a multistep process. After  photoexcitation, the `bright' state undergoes ultrafast internal conversion to the `dark' state. The `dark' state (often labeled the $S_1$ or $2A_g$ state) is a linear combination of an entangled singlet triplet-pair and an odd-parity charge-transfer exciton. By building a minimal model of the triplet-pair component of the `dark' state and using the quantum Liouville equation, this paper has described the subsequent fate of the triplet-pair in a carotenoid dimer. Our key findings are summarized in Section \ref{Se:5.3}.

The theory developed in this paper is predicated on the assumption that in carotenoids and polyenes, unlike in acenes, the initial step of forming an entangled triplet-pair from the photoexcited electron-hole pair is an intrachain process. However, as just indicated, the `dark' state has some charge-transfer character. Moreover, the initial photoexcited electron-hole pair does not evolve completely into the `dark' state, but retains some Frenkel-exciton `bright' state character\cite{Manawadu2022,Manawadu2023a}. Thus, a complete theory of singlet fission in carotenoid systems should include the singlet electron-hole components of the initial state to determine whether there is also bimolecular triplet-pair formation.

This paper has also focussed on singlet fission from the lowest member of the `$2A_g$' family of states, for which intrachain singlet fission is strongly endothermic. However, theory\cite{Valentine20,Manawadu2022,Manawadu2023a} and experiment\cite{Frank1997,Kosumi2006} both indicate that internal conversion to the $2^1A_g$ state may occur via intermediate states, i.e., via higher energy members of the `$2A_g$' family of states from which intrachain singlet fission is potentially exothermic\cite{Valentine20}. Again, this possibility needs to be investigated in further work.

Using a model Hamiltonian with adjustable parameters, we have been able to explore the `parameter space' of singlet fission in carotenoid dimers, thus understanding the kinetics and predicting the equilibrated yields. However, in order to make concrete predictions to explain experimental observations we will need to include additional interactions, e.g., dipolar and spin-orbit coupling, as well as to derive \emph{ab initio} parameters for realistic carotenoid conformations. In conclusion, this paper has presented a first step to theoretically model the rich and fascinating physics of singlet fission in carotenoid systems, but there is clearly much more theoretical and computational modelling to do.


\begin{acknowledgments}
We thank Claudia Tait and Max Marcus for helpful discussions.
\end{acknowledgments}



\appendix

\section{Heisenberg Model of Triplet-Pair States}\label{Se:A2}

The low-energy spin-physics of carotenoids and polyenes is approximately described by the spin-1/2 dimerized Heisenberg antiferromagnet\cite{Book}. Its Hamiltonian is,
\begin{equation}\label{}
  \hat{H} = J_d \sum_{n \in odd} \hat{\textbf{S}}_n \cdot \hat{{\textbf{S}}}_{n+1} +  J_s \sum_{n \in even} \hat{\textbf{S}}_n \cdot \hat{\textbf{S}}_{n+1},
\end{equation}
where $n$ labels  a p$_z$ orbital on the $n$th C-atom and
$\hat{{\textbf{S}}}_{n}$ is the spin-1/2 operator acting on the electron in that orbital.

$J_d$ and $J_s$ are the superexchange parameters for spin-1/2 electrons across the double and single bonds, respectively. Their origin lies in electron transfer between orbitals via a virtual, higher energy  charge-transfer state. Thus, they are proportional to $\beta^2$, where $\beta$ is the one-electron transfer integral between the p$_z$ orbitals.

We assume that the groundstate of a carotenoid is a product of singlet dimers located on each double bond. A single triplet excitation corresponds to exciting one of these singlet dimers into a triplet, labelled $i$, with an excitation energy, $E_T = J_d$. Similarly, a triplet-pair  corresponds to two excited singlet dimers, labelled $i$ and $j$, with an excitation energy $2J_d$. Two triplets can be coupled to form a singlet, triplet or quintet spin-eigenstate. The $S_z=0$ components of these states are given in eqn (2-4).

\begin{figure}
\includegraphics[width=1.0\linewidth]{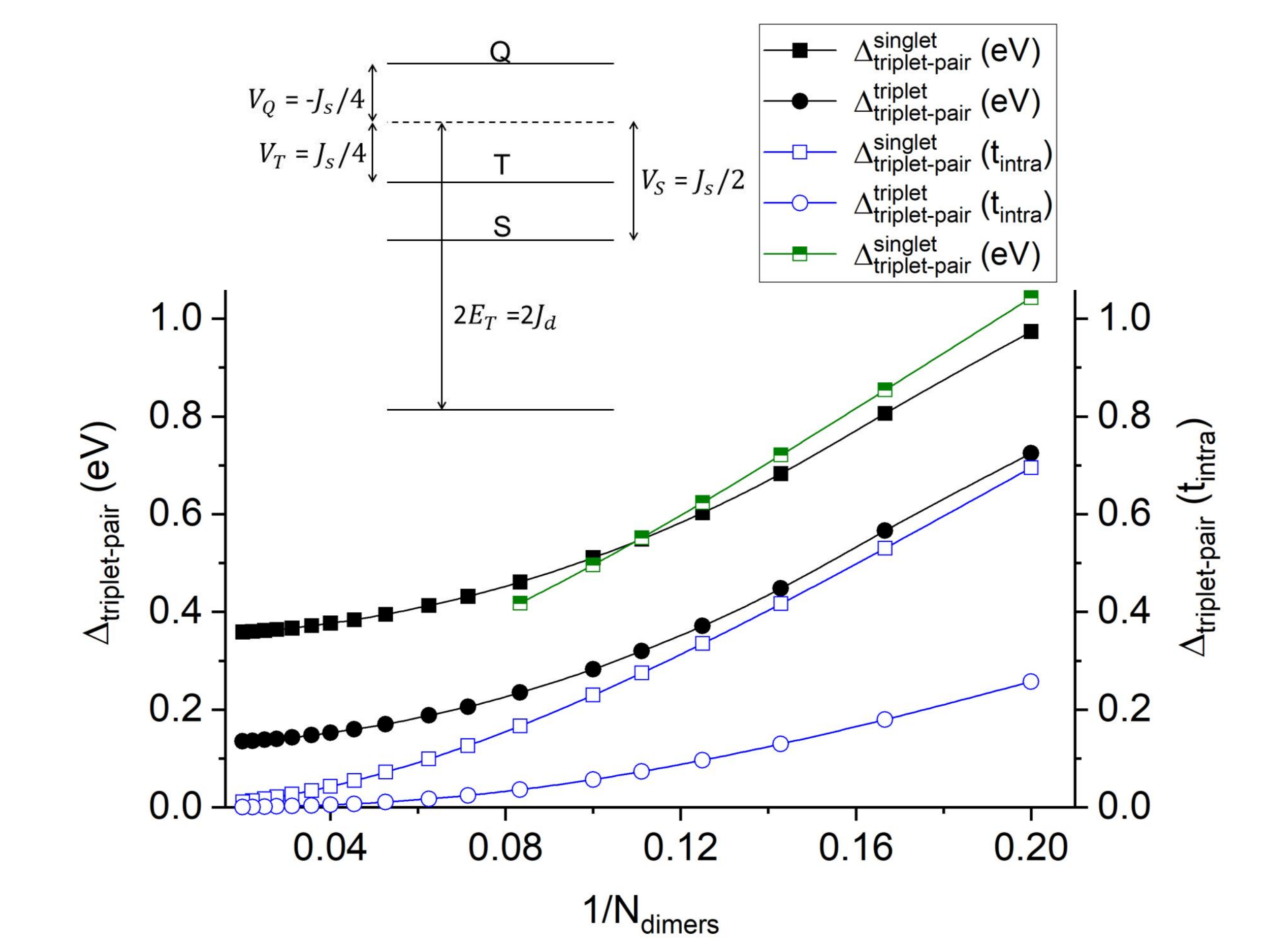}
\caption{Binding energies of the singlet and triplet triplet-pair states on a linear chain of $N_{\textrm{dimers}}$. The results in blue with open symbols are obtained from the one-chain Hamiltonian given in eqn (\ref{Eq:401}), using parameters derived from the antiferromagnetic Heisenberg model (AFHM), i.e., $V_S = 2$\tintra\ and $V_T = $\tintra\ (right ordinate).
The results in black with filled symbols are obtained from the  triplet-pair-charge-transfer-exciton basis model of ref\cite{Barford2022c} (left ordinate).
These latter results can be reproduced to a good approximation by eqn (\ref{Eq:401}) when $V_S = 2.8$\tintra\ and \tintra$ = 0.88$ eV, as shown by the half-full square symbols in green (left ordinate).
The inset shows the singlet, triplet and quintet triplet-pair energy levels on two dimers, with the spin-1 exchange-interaction parameters determined via the AFHM.}
\label{Fi:16}
\end{figure}

Rewriting $\hat{\textbf{S}}_n \cdot \hat{\textbf{S}}_{n+1}$ as\\ $\left(\hat{S}_n^z  \hat{S}_{n+1}^z +(\hat{S}_n^+  \hat{S}_{n+1}^- + \hat{S}_n^-  \hat{S}_{n+1}^+)/2\right)$ and using the triplet-pair wavefunctions (eqn (2-4)), it is easy to show that:
\begin{itemize}
\item{Triplets  hop across the single-bond between neighboring dimers with a hopping matrix element, \tintra\ $ = -J_s/4$.}
\item{A pair of triplets in an overall singlet eigenstate experiences a nearest neighbor-dimer \emph{attraction}, $V_S = J_s/2$.}
\item{A pair of triplets in an overall triplet eigenstate experiences a nearest neighbor-dimer \emph{attraction}, $V_T = J_s/4$.}
\item{A pair of triplets in an overall quintet eigenstate experiences a nearest neighbor-dimer \emph{repulsion}, $V_Q = -J_s/4$.}
\end{itemize}
The energy levels for a triplet-pair occupying a pair of dimers are shown in the inset of Fig.\ \ref{Fi:16}. (These relative energy levels may also be obtained by diagonalizing $J \hat{\textbf{S}}_1^{(1)} \cdot \hat{{\textbf{S}}}_{2}^{(1)}$, where $\hat{\textbf{S}}^{(1)}$ is the  spin-1 operator and $J = J_s/4$.\cite{Kollmar1993})

For our purposes, we are interested in a pair of itinerant, interacting triplets on a chain of dimers. This two-particle scattering problem has been solved\cite{Mattis1988,Gallagher1997,Gebhard1997}. The key result is discussed in Section \ref{Se:3.1}, namely that for an infinite chain a single bound state exists when $V > 2$\tintra. Thus, the parameters derived from the spin-1/2 Heisenberg antiferromagnet do not predict bound triplet-pairs for an infinite chain, because $V_S = J_s/2= 2$\tintra.

Since the  triplet-pair component of the singlet `dark' state of carotenoids and polyenes is strongly bound, we learn from this analysis that the spin-1/2  antiferromagnet Heisenberg model does not fully describe  this state. Instead, the `dark' state of carotenoids is more correctly described as a linear combination of a singlet triplet-pair and a \emph{real} charge-transfer state. As shown in ref\cite{Barford2022c}, it is the hybridization between these two components that causes a stronger attractive interaction (or exchange coupling) between the triplets than that predicted by the pure spin-1/2 Heisenberg model.
Figure  \ref{Fi:16} shows the binding energy of the singlet and triplet triplet-pair states derived from both the Heisenberg model and the model of ref\cite{Barford2022c}.

Unfortunately, although still  a reduced basis for the many-body problem, the triplet-pair-charge-transfer-exciton basis\cite{Barford2022c} (which itself is a reduced basis version of the exciton-basis valence-bond theory\cite{Chandross1999}) is too large to perform quantum Liouville simulations for realistic chain lengths.
It is therefore expedient to retain the minimal valence-bond basis, i.e., of a pair of triplets moving in a chain of singlet dimers, but with parameters derived by fitting to the model of ref\cite{Barford2022c}. We find that an nearest-neighbor attraction of $V_S = 2.8$\tintra\ with \tintra $= 0.88$ eV reproduces the binding energies derived from the model of ref\cite{Barford2022c} shown in Fig.\ \ref{Fi:16}.



\bibliography{references}

\end{document}